\pdfoutput=1
\newcommand*{\ATLASLATEXPATH}{}
\documentclass[PAPER, texlive=2016, cernpreprint, UKenglish]{\ATLASLATEXPATH atlasdoc}
\usepackage[subcaption]{\ATLASLATEXPATH atlaspackage}
\usepackage[backref=false]{\ATLASLATEXPATH atlasbiblatex}
 
\usepackage[bsm]{\ATLASLATEXPATH atlasphysics}
 
\graphicspath{{logos/}{figures/}}
 
\usepackage{ANA-SUSY-2018-09-PAPER-defs}
\usepackage{multirow}
 
\LEcontact{Bob Kowalewski bob.kowalewski@cern.ch}

\AtlasTitle{Search for squarks and gluinos in final states with same-sign leptons and jets using 139 fb$^{-1}$ of data collected with the ATLAS detector}
 
\AtlasVersion{3.0}
 
\AtlasAbstract{
A search for supersymmetric partners of gluons and quarks is presented,
involving signatures with jets and either two isolated leptons (electrons or muons) with the same electric charge, or at least three isolated leptons.
A data sample of proton--proton collisions at $\sqrt{s}$ = 13 TeV recorded with the ATLAS detector at the Large Hadron Collider between 2015 and 2018,
corresponding to a total integrated luminosity of 139 fb$^{-1}$, is used for the search.
No significant excess over the Standard Model expectation is observed.
The results are interpreted in simplified supersymmetric models
featuring both $R$-parity conservation and $R$-parity violation,
raising the exclusion limits beyond those of previous ATLAS searches
to 1600 GeV for gluino masses and 750 GeV for bottom and top squark masses in these scenarios.
}
 
\author{The ATLAS Collaboration}
 
\AtlasRefCode{SUSY-2018-09}
 
\PreprintIdNumber{CERN-EP-2019-161}

\AtlasCoverSupportingNote{Supporting note}{https://cds.cern.ch/record/2648515}
 
\AtlasCoverCommentsDeadline{2020, March 29}
 
\AtlasCoverAnalysisTeam{N.~Abraham, O.~Ducu, Y.~Liu, J.~Maurer, D.~Paredes~Hernandez, P.~Tornamb\'e, A.~De~Santo, X.~Zhuang}
 
\AtlasCoverEdBoardMember{J.~Boyd, F.~Cardillo, G.~Polesello~(chair)}
 
\AtlasCoverEgroupEditors{atlas-SUSY-2018-09-editors@cern.ch}
 
\AtlasCoverEgroupEdBoard{atlas-SUSY-2018-09-editorial-board@cern.ch}
 
\AtlasJournal{JHEP}
 
\AtlasJournalRef{JHEP 06 (2020) 46}
 
\AtlasDOI{10.1007/JHEP06(2020)046}
\hypersetup{pdftitle={ATLAS document},pdfauthor={The ATLAS Collaboration}}
 
\begin{document}
 
\maketitle
 
\section{Introduction}
\label{sec:intro}
Experimental searches for manifestations of physics beyond the Standard Model (BSM physics) at hadron colliders
have long exploited the signature of final states comprising a pair of isolated light leptons (electrons, muons)
with the same electric charge (`same-sign leptons').
In the Standard Model (SM), production of prompt same-sign lepton pairs from weak-boson decays is rare.
In the context of $\sqrt{s}=\SI{13}{TeV}$ $pp$ collisions,
the inclusive cross-section is of the order of \SI{1}{pb}~\cite{Alwall:2014hca, Tanabashi:2018aaa},
thus suppressed by more than three orders of magnitude
relative to the production of opposite-sign lepton pairs.
By constrast, in many scenarios heavy BSM particles, which may be produced in proton--proton ($pp$) collisions,
decay into multiple massive SM bosons or top quarks.
The subsequent decays of these heavy SM particles into same-sign leptons and jets may then occur with significant branching ratios.
Pair production of heavy BSM Majorana fermions can be another abundant source of events with same-sign leptons~\cite{Barnett:1993ea}.
 
At the Large Hadron Collider (LHC)~\cite{Evans:2008zzb},
signatures with same-sign prompt leptons have been used by the ATLAS~\cite{PERF-2007-01} and CMS~\cite{CMS-TDR-08-001} experiments
to explore the landscape of possible SM extensions and their phenomenology.
Among these proposed extensions,
supersymmetry (SUSY)~\cite{Golfand:1971iw,Volkov:1973ix,Wess:1974tw,Wess:1974jb,Ferrara:1974pu,Salam:1974ig}
stands out as a particularly compelling framework.
It was shown~\cite{Sakai:1981gr,Dimopoulos:1981yj,Ibanez:1981yh,Dimopoulos:1981zb}
to favourably impact the scale evolution of perturbative gauge couplings needed for
the unification of strong and electroweak interactions,
and can address the SM gauge hierarchy problem.
In its minimal realisation, the MSSM~\cite{Fayet:1976et,Fayet:1977yc},
each fundamental SM fermion is associated with a pair of new scalar partners
--- in the case of quarks $q$, the squarks $\tilde q_\text{L}$ and $\tilde q_\text{R}$.
Similarly, each SM bosonic degree of freedom is partnered with a new fermion.
Mixing between the partners of SM electroweak and Higgs bosons\footnote{The Higgs sector is also enriched by the presence of an additional complex doublet.}
results in four massive Majorana fermions and two massive charged fermions
(neutralinos $\tilde\chi_1^0$ to $\tilde\chi_4^0$ and charginos $\tilde\chi_1^\pm$ and $\tilde\chi_2^\pm$, indexed by increasing mass).
The gluinos $\tilde g$, partners of the SM gluons, do not mix due to their colour charge.
 
SUSY can provide a massive dark-matter candidate~\cite{Goldberg:1983nd,Ellis:1983ew},
the lightest supersymmetric particle (LSP), if an additional ad hoc discrete symmetry, called $R$-parity~\cite{Farrar:1978xj},
is invoked.
When this symmetry is conserved,
supersymmetric partners can only be produced in pairs and decay into the LSP and SM particles,
possibly in several steps via superpartners of intermediate masses.
The LSP, stable and weakly interacting, escapes the detector,
leaving a striking experimental signature of large missing transverse momentum.
When $R$-parity is not conserved, the final states contain only SM particles;
decay channels for squarks include e.g.\ $\tilde q_i\to q_j q_k$ or $\tilde q_i\to q_j \ell_k$,
if the corresponding coupling strengths~\cite{Martin:1997ns} $\lambda''_{ijk}$ or $\lambda'_{ijk}$ are non-zero.
 
Naturalness arguments~\cite{Barbieri:1987fn,deCarlos:1993yy}
suggest that the top squark mass may not exceed $\approx\SI{1}{TeV}$~\cite{Inoue:1982pi,Ellis:1983ed}.
Significant mixing between the scalar top partners $\tilde t_\text{L}$ and $\tilde t_\text{R}$,
enhanced relative to other quark flavours,
can also lower the mass of the lightest eigenstate $\stopone$ below that of other squarks.
These constraints indirectly affect gluinos and bottom squark masses as well.
Gluinos and third-generation squarks may therefore be among the superpartners with low mass
and copiously produced at the LHC.
Typical pair-production cross-sections~\cite{Beenakker:2016lwe} for interesting scenarios in the context of this paper
are \SI{9}{fb} for a \SI{1.6}{TeV} gluino mass,
or \SI{33}{fb} for the lightest top $\tilde t_1$ or bottom $\tilde b_1$ squark mass of \SI{800}{GeV}.
 
\begin{figure}[t!]
\begin{subfigure}{0.24\textwidth}
\includegraphics[width=\textwidth]{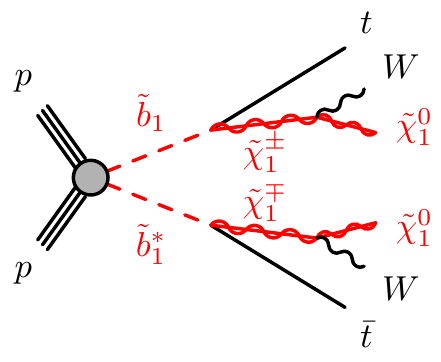}
\subcaption{}\label{fig:sbottom-tWN1}
\end{subfigure}
\begin{subfigure}{0.24\textwidth}
\includegraphics[width=\textwidth]{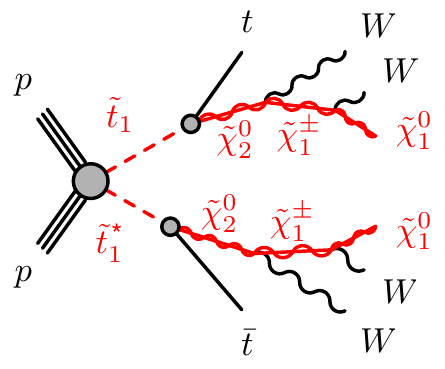}
\subcaption{}\label{fig:stop-tWWN1}
\end{subfigure}
\begin{subfigure}{0.24\textwidth}
\includegraphics[width=\textwidth]{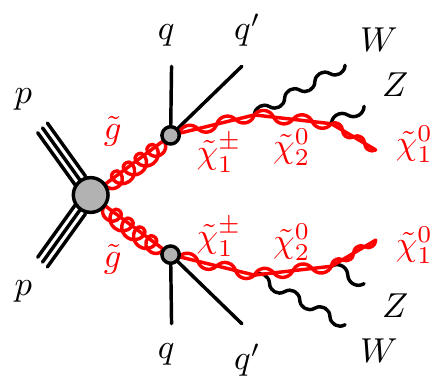}
\subcaption{}\label{fig:gluino-qqWZN1}
\end{subfigure}
\begin{subfigure}{0.24\textwidth}
\includegraphics[width=\textwidth]{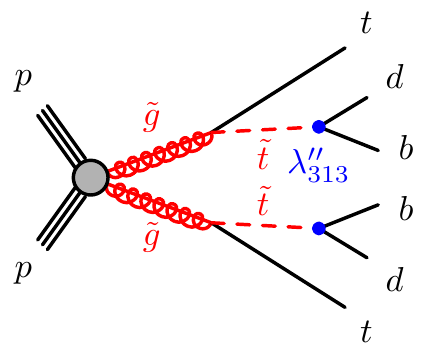}
\subcaption{}\label{fig:gluino-tbd}
\end{subfigure}
\caption{
Examples of processes allowed in the MSSM,
involving the pair production and cascade decays of squarks and gluinos into final states with leptons and jets.
}
\label{fig:benchmarks}
\end{figure}

This paper presents a search for gluinos and squarks
in final states with two same-sign leptons and jets.
The events may include additional leptons.
In addition, large missing transverse momentum is required in the case of $R$-parity-conserving models.
The event selection also relies on the number of $b$-tagged jets.
Signal regions (SRs) are built (Section~\ref{sec:selection}) from a set of requirements on the kinematic properties of the selected events,
in order to isolate the signature of supersymmetric processes from SM backgrounds.
The latter are estimated with Monte Carlo simulation for processes such as $t\bar tV$ or $VV$ ($V=W,Z$) leading to prompt same-sign leptons (Section~\ref{sec:prompt_bkg}),
while sources of same-sign leptons arising from jets misidentified as leptons or non-prompt leptons from decays of hadrons,
as well as other reducible backgrounds,
are estimated with data (Section~\ref{sec:reducible_bkg}).
Event yields in data are then compared with the estimated contributions from SM processes.
The results are presented in Section~\ref{sec:results} for \SI{139}{fb^{-1}} of \SI{13}{TeV} $pp$ collision data recorded by the ATLAS experiment.
They are interpreted in terms of exclusion limits (Section~\ref{sec:limits}) on the parameters of four benchmark supersymmetric signal scenarios,
which are shown in Figure~\ref{fig:benchmarks}.
 
A similar, earlier analysis, realised on a subset of the data for these results, was reported in Ref.~\cite{SUSY-2016-14}
and found no deviation from SM expectations.
Searches based on these event topologies were also performed in the same context with the CMS experiment with the same outcome~\cite{CMS-SUS-16-035,CMS-SUS-16-041}.

\section{ATLAS detector}
\label{sec:atlas}
The ATLAS experiment~\cite{PERF-2007-01} at the LHC is a multipurpose particle detector
with a forward--backward symmetric cylindrical geometry and a near $4\pi$ coverage in
solid angle.\footnote{
ATLAS uses a right-handed coordinate system with its origin at the nominal interaction point (IP)
in the centre of the detector and the $z$-axis along the beam pipe.
The $x$-axis points from the IP to the centre of the LHC ring,
and the $y$-axis points upwards.
Cylindrical coordinates $r,\phi$ are used in the transverse plane,
$\phi$ being the azimuthal angle around the $z$-axis.
The pseudorapidity is defined in terms of the polar angle $\theta$ as $\eta = -\ln \tan(\theta/2)$.
The rapidity is defined relative to the beam axis as a function of the velocity $\beta$: $y=0.5\times\ln\{(1+\beta\cos\theta)/(1-\beta\cos\theta)\}$.
The magnitude of the momentum in the plane transverse to the beam axis is denoted by \pt.
}
It consists of an inner tracking detector (ID) surrounded by a thin superconducting solenoid
providing a \SI{2}{\tesla} axial magnetic field, electromagnetic (EM) and hadron calorimeters, and a muon spectrometer (MS).
The ID covers the pseudorapidity range \(|\eta| < 2.5\).
It consists of silicon pixel, silicon microstrip, and transition radiation tracking detectors,
completed by the insertable B-layer (IBL) installed before \RunTwo~\cite{ATLAS-TDR-2010-19,PIX-2018-001}.
Lead/liquid-argon (LAr) sampling calorimeters provide EM energy measurements
with high granularity.
A steel/scintillator-tile hadron calorimeter covers the central pseudorapidity range $\abseta < 1.7$.
The endcap and forward regions are instrumented with LAr calorimeters
for EM and hadronic energy measurements up to $\abseta=4.9$.
The MS surrounds the calorimeters and is based on
three large air-core toroidal superconducting magnets with eight coils each.
The field integral of the toroids ranges between \num{2.0} and \SI{6.0}{\tesla}$\cdot$m
across most of the detector.
The MS includes a system of precision tracking chambers and fast detectors for triggering.
A two-level trigger system~\cite{TRIG-2016-01} is used to select events.
The first-level trigger is implemented in hardware and uses a subset of the detector information
to reduce the accepted rate to at most \SI{100}{\kHz}.
This is followed by a software-based trigger that
reduces the accepted event rate to \SI{1}{\kHz} on average
depending on the data-taking conditions.
 
\section{Event reconstruction}
\label{sec:reconstruction}
The analysis is performed on a set of $pp$ collision data
recorded by the ATLAS detector between 2015 and 2018.
In this period, the LHC delivered colliding beams
with a peak instantaneous luminosity up to $L=\SI{2.1e34}{cm^{-2}s^{-1}}$ achieved in 2018,
and an average number of $pp$ interactions per bunch crossing (`pile-up') of \num{33.7}.
After requirements on the stability of the beams,
the operational status of all ATLAS detector components,
and the quality of the recorded data,
the total integrated luminosity of the dataset corresponds to \SI{139}{fb^{-1}} with an uncertainty of $1.7\%$.
It is obtained~\cite{ATLAS-CONF-2019-021} using the LUCID-2 detector \cite{LUCID2} for the primary luminosity measurements.
 
Proton--proton interaction vertices are reconstructed from charged-particle tracks in the ID with $\pt>\SI{500}{MeV}$~\cite{ATL-PHYS-PUB-2015-026,PERF-2015-01}.
The presence of at least one such vertex with a minimum of two associated tracks is required,
and the primary vertex is chosen as the vertex with the largest sum of \pTsq\ of associated tracks.
 
The \antikt algorithm~\cite{Cacciari:2008gp} with radius parameter $R=0.4$
implemented in the \texttt{FastJet} library~\cite{Fastjet}
is used to reconstruct jets up to $\abseta=4.9$,
relying on topological energy clusters in the calorimeter~\cite{PERF-2014-07} at the EM scale.
Jets are then calibrated as described in Ref.~\cite{PERF-2016-04}.
Only jets with $\pt>\SI{20}{GeV}$ are further considered.
Events are vetoed when containing jets induced by calorimeter noise or non-collision background,
according to criteria similar to those described in Ref.~\cite{ATLAS-CONF-2015-029}.
As decay products of heavy particles tend to be more central, this analysis only considers jets with $\abseta<2.8$ in multiplicity-based requirements.
An additional discriminant referred to as the \texttt{Jet Vertex Tagger} (\texttt{JVT})
is used to exclude
jets produced in pile-up processes~\cite{PERF-2014-03},
based on classifying the tracks associated with the jet as pointing or not pointing to the primary vertex.
 
Jets containing $b$-flavoured hadrons are identified in the region $\abseta<2.5$ by the MV2c10 \btag algorithm~\cite{Aad:2019aic},
which makes use of the impact parameters of tracks associated with the candidate jet,
the position of reconstructed secondary vertices and their consistency with the decay chains of such hadrons.
For the working point chosen for this analysis,
such jets are tagged with an efficiency of $70\%$ in simulated $t\bar t$ events,
with mis-tag rates of $9\%$ and $0.3\%$ for jets initiated by charm quarks or light quarks/gluons, respectively.
 
Baseline muon candidates are reconstructed~\cite{PERF-2015-10} in the region $\abseta<2.5$ from MS tracks matching ID tracks.
The analysis only considers muons with $\pt>\SI{10}{GeV}$
satisfying the set of requirements on the quality of the tracks which is defined as \texttt{Medium} in Ref.~\cite{PERF-2015-10}.
The longitudinal impact parameter $z_0$ of the muon track
must satisfy $|z_0\sin\theta|<\SI{0.5}{mm}$.
Signal muons are defined as the baseline candidates sufficiently distant from jets (see below) and other leptons,
which satisfy further requirements:
the transverse impact parameter $d_0$ of the track must be sufficiently small relative to its uncertainty from the track reconstruction, $|d_0|<3\sigma(d_0)$,
and the candidate must satisfy a track-based isolation criterion.
The latter requires the summed scalar $\pt$ of nearby ID tracks not to exceed 6\% of the muon \pt,
for selected tracks in a \pt-dependent $\Delta R_\eta = \sqrt{(\Delta\eta)^2+(\Delta\phi)^2}$ cone of maximal size 0.3 around the muon, excluding its own track,
similarly to the isolation variables defined in Ref.~\cite{PERF-2015-10};
these tracks must be associated with the primary vertex to limit sensitivity to pile-up.
 
Baseline electron candidates are reconstructed~\cite{EGAM-2018-01} from energy depositions in the EM calorimeter
matched to an ID track and are required to have $\pt>\SI{10}{GeV}$ and $\abseta<2.47$,
excluding the transition region $1.37<\abseta<1.52$ between the barrel and endcap EM calorimeters.
They must satisfy the \texttt{LooseAndBLayerLLH} identification discriminant defined in Ref.~\cite{EGAM-2018-01},
as well as requirements on the track impact parameters $|z_0\sin\theta|<\SI{0.5}{mm}$ and $|d_0|<5\sigma(d_0)$.
Signal electrons, which must be distant from jets and other leptons,
are required to satisfy the tighter \texttt{MediumLLH} identification
and \texttt{FCTight} isolation requirements defined in Ref.~\cite{EGAM-2018-01}.
The latter are similar to the muon isolation requirement, with a maximal cone size of 0.2,
but with an additional calorimeter-based isolation requirement using nearby topological clusters instead of tracks.
Only signal electrons with $\abseta<2.0$ are considered in order to reduce the rate of electrons with wrongly reconstructed charge (`charge-flip');
the latter are further rejected by the application of the \texttt{ECIDS} discriminant described in Ref.~\cite{EGAM-2018-01},
which exploits further information related to the electron track reconstruction and its compatibility with the primary vertex and the electron cluster.
 
The missing transverse momentum (whose magnitude is denoted \met) is defined
as the negative vector sum of the transverse momenta of all identified objects
(baseline electrons, photons~\cite{EGAM-2018-01}, baseline muons and jets) and an additional soft term.
The soft term is constructed from all tracks associated with the primary vertex but not with any physics object.
In this way, the \met\ is adjusted for the best calibration of the jets and the other identified physics objects
listed above, while maintaining approximate pile-up independence in the soft term~\cite{PERF-2016-07,ATLAS-CONF-2018-023}.
Overlaps between objects in the \met\ calculation are resolved as described in Ref.~\cite{PERF-2016-07}.
 
To exclude non-prompt leptons produced inside jets,
baseline leptons close to jets are discarded
according to the angular distance $\Delta R=\sqrt{(\Delta y)^2+(\Delta\phi)^2}$
between the two reconstructed objects.
A requirement of $\Delta R>\operatorname{min}\{0.4, 0.1+\SI{9.6}{GeV}/\pt(\ell)\}$ is used.
 
\section{Event selection}
\label{sec:selection}
Events are selected if they contain at least two signal leptons with $\pt>\SI{20}{GeV}$.
In addition, there must be at least one pair of leptons with identical electric charges
among the ensemble of signal leptons with $\pt>\SI{10}{GeV}$.
 
Data events were recorded via a combination of triggers
based on the presence of missing transverse momentum
or pairs of leptons~\cite{ATL-DAQ-PUB-2016-001,ATL-DAQ-PUB-2017-001,ATL-DAQ-PUB-2018-002,TRIG-2018-05}.
For events with $\met<\SI{250}{GeV}$,
only lepton-based triggers without isolation requirements are used,
with lepton \pt\ thresholds which
vary over the data collected in \RunTwo
up to a maximum of \SI{24}{GeV} for triggers requiring two electrons,
$\SI{22}{GeV}$ for the leading-\pt\ muon in triggers requiring two muons,
and $\SI{17}{GeV}$ (\SI{14}{GeV}) for the electron (muon) in mixed dilepton triggers.
For events with $\met>\SI{250}{GeV}$, triggers based on \met\ are also used.
For events that are only accepted by lepton triggers with \pt\ requirements above \SI{20}{GeV},
the analysis-level lepton \pt\ requirement is raised to be \SI{1}{GeV} above the trigger threshold.
This results in a relative reduction of the total fiducial acceptance by at most 2\%
for the benchmark signal scenarios of Figure~\ref{fig:benchmarks}.
For signal events selected in the SRs presented below,
the trigger efficiency is above 95\% for $R$-parity-conserving models,
and above 93\% otherwise.
For signal events with $\met>\SI{250}{GeV}$, the trigger efficiency is above 99\%.

Five SRs are built to isolate signatures of hypothetical supersymmetric signal processes
from backgrounds;
their definitions are summarised in Table~\ref{tab:SR-definitions}.
They rely on the multiplicities of different reconstructed objects
such as
the number of leptons $n_\ell$ and their relative electric charges,
the number of jets $n_j$ with $\pt>25$ or \SI{40}{GeV},
and the number of \btagged jets $n_b$ with $\pt>\SI{20}{GeV}$.
Several kinematic variables are also used:
the effective mass $\meff$ consisting of the scalar $\pt$ sum of all jets and leptons added to \met,
the \met\ itself and its ratio to \meff,
and the invariant mass of same-sign electron pairs, $m_{e^\pm e^\pm}$.
The latter helps to reduce the backgrounds featuring a $Z\to e^+e^-$ decay where the charge of one electron is mismeasured.
The SR requirements were chosen loosely so as to provide sensitivity to
non-excluded regions of the parameter space for the processes illustrated in Figure~\ref{fig:benchmarks},
while preserving sensitivity to other SUSY processes with possibly different final states,
as in Table~\ref{tab:SR-definitions}.
 
\begin{table}[t!]
\caption{
Definition of the signal regions used by the analysis,
based on the variables defined in Section~\ref{sec:selection}.
The last column provides examples of SUSY processes
which may contribute to these signal regions.
}
\label{tab:SR-definitions}
 
\def\arraystretch{1.1}
\resizebox{\textwidth}{!}{
\begin{tabular}{|c|c|c|c|c|c|c|c|}
\hline
\hline
SR &  $n_{\ell}$   & $n_{b}$    & $n_{j}$  & \met\ [GeV] & \meff\ [GeV]  & $\met/\meff$ & SUSY\\
\hline\hline
\texttt{Rpv2L} & $\geq 2$ ($\ell^\pm\ell^\pm$) & $\geq 0$ & $\geq 6$ ($\pt>\SI{40}{GeV}$) & $-$ & $>2600$ & $-$ & $\gluino\to t\tilde t_1^*,\;\tilde t_1^*\to qq'$ ($\lambda''\neq 0$) \\
&  &  & & & & &  $\gluino\to t\bar t\ninoone,\;\ninoone\to 3q$ $(\lambda''\neq 0)$      \\
&  &  & & & & & $\gluino\to q\bar q\ninoone,\;\ninoone\to qq'\ell$ $(\lambda'\neq 0)$ \\
\hline\hline
\texttt{Rpc2L0b}      & $\geq 2$ ($\ell^\pm\ell^\pm$) & $=0$ & $\geq 6$ ($\pt>\SI{40}{GeV}$)& $>200$ & $>1000$ & $>0.2$  &  $\gluino\to q\bar q' WZ\ninoone$ \\
\hline\hline
\texttt{Rpc2L1b} & $\geq 2$ ($\ell^\pm\ell^\pm$) &$\geq 1$ & $\geq 6$ ($\pt>\SI{40}{GeV}$) & $-$ & $-$ & $>0.25$ &  $\sbottomone\to tW\ninoone$ \\
\hline\hline
\texttt{Rpc2L2b} & $\geq 2$ ($\ell^\pm\ell^\pm$) &$\geq 2$ & $\geq 6$ ($\pt>\SI{25}{GeV}$) & $>300$ & $>1400$ & $>0.14$ &  $\sbottomone\to tW\ninoone$\\
& & & & & & & $\gluino\to \ttbar\ninoone$ \\
\hline\hline
\texttt{Rpc3LSS1b}  & $\geq 3$ ($\ell^\pm\ell^\pm\ell^\pm$) & $\geq 1$ &\multicolumn{3}{c|}{no cut but veto $\SI{81}{GeV}<m_{e^\pm e^\pm}<\SI{101}{GeV}$}         & $>0.14$     & $\stopone \to tW^{\pm}(W^*)\ninoone$  \\
\hline\hline
\end{tabular}
}
\end{table}
 
The SR \texttt{Rpv2L} targets gluino pair production in $R$-parity-violating scenarios, hence without any \met\ requirement.
It is inclusive in terms of \btagged jets
to be sensitive to various decay modes of gluinos leading to final states with leptons and jets,
such as the scenario illustrated in Figure~\ref{fig:gluino-tbd}
or the few other examples mentioned in Table~\ref{tab:SR-definitions}.
In this SR, a tight requirement on the effective mass $\meff>\SI{2.6}{TeV}$ is used to reduce SM backgrounds.
 
The SR \texttt{Rpc2L0b} provides sensitivity to $R$-parity-conserving scenarios not involving third-generation squarks,
as in Figure~\ref{fig:gluino-qqWZN1},
which are less likely to contain bottom quarks in the final state.
A veto on \btagged jets is imposed in order to reduce SM backgrounds with top quarks.
Requiring a large number of jets strongly reduces the level of $WZ$ and other multiboson backgrounds.
 
The SRs \texttt{Rpc2L1b} and \texttt{Rpc2L2b}
provide sensitivity to scenarios involving third-generation squarks,
such as $\sbottomone\to t\chinoonem$ with a subsequent $\chinoonepm\to W^\pm\ninoone$ decay as in Figure~\ref{fig:sbottom-tWN1}.
\texttt{Rpc2L2b} uses tighter requirements on \met\ and \meff\ than \texttt{Rpc2L1b}
in order to complement it at low $\ninoone$ mass,
as well as to provide good sensitivity to scenarios with heavier superpartners such as pair-produced gluinos decaying via $\gluino\to t\bar t\ninoone$.
 
Finally, the SR \texttt{Rpc3LSS1b} probes scenarios with long decay chains but compressed mass spectra
leading to final states with softer decay products, such as the $\stopone\to t\ninotwo\to tW(W^*)\ninoone$ cascade decay shown in Figure~\ref{fig:stop-tWWN1} and  proposed in Ref.~\cite{Huang:2015fba}.
This SR selects events with at least three leptons of identical charge,
leading to a huge reduction of the expected SM background yields.
Loose requirements on the $\met/\meff$ ratio and the presence of at least one \btagged jet,
as well as the rejection of events containing any pair of same-sign electrons
with $m_{e^\pm e^\pm}$ close to the $Z$ boson mass,
help to diminish the residual reducible background to low levels.

A simple cut-and-count analysis is performed in each SR.
The number of events in data is reported in Section~\ref{sec:results}
together with the expected contributions from SM processes and the reducible background,
the estimations of which are described in the following sections.
 
\section{Standard Model backgrounds}
\label{sec:prompt_bkg}
Major contributions from SM processes to the SRs
arise from $WZ$+jets (with minor contributions from $ZZ$ and $W^\pm W^\pm jj$\footnote{
This process corresponds to the production of two same-sign $W$ bosons~\cite{ATL-PHYS-PUB-2019-004}
which at the lowest order of the perturbative expansion are accompanied by two forward jets.
}),
$t\bar tW$ and $t\bar tZ$.
The summed contributions of other processes involving associated production
of top quarks and massive bosons,
with smaller production cross-sections,
can also amount to significant fractions of the expected SR event yields.
SRs with at least one \btagged jet are populated mainly by processes involving top quarks,
while multiboson processes dominate in regions vetoing \bjets.
In the case of the \texttt{Rpc3LSS1b} SR,
only processes such as $WZZ$, $ZZZ$, $t\bar tWZ$ and $VH/t\bar tH$ where the Higgs boson $H$ decays via $H\to 4\ell$
are genuine sources of events with three same-sign prompt leptons.
 
\begin{table}[t!]
\caption{
List of Monte Carlo event generators and their settings
used to predict the contributions from SM processes
to the various regions of interest in the analysis.
When no reference is provided for the cross-section normalisation,
the one computed by the generator is used.
The LO and NLO acronyms are defined in Section~\ref{sec:prompt_bkg}.
}
\label{tab:bkg-generators}
 
\scriptsize
\resizebox{\textwidth}{!}
{
\begin{tabular}{|l|l|l|l|l|l|l|}
\hline\hline
Physics process    & Event generator & Computation & Parton shower & Cross-section & PDF set & Set of tuned \\
&	             & order       &               & normalisation &         & parameters  \\
\hline
\hline
$t\bar{t}W$~\cite{ATL-PHYS-PUB-2016-005}  &  MG5\_\AMCatNLOV{2.3.3}~\cite{Alwall:2014hca}  &  NLO  &  \PYTHIAV{8.210}~\cite{Sjostrand:2014zea}  &  NLO~\cite{deFlorian:2016spz}  &  NNPDF2.3LO~\cite{Ball:2012cx}  &  A14~\cite{ATL-PHYS-PUB-2014-021}    \\
 
$t\bar{t}Z/\gamma^{*}$~\cite{ATL-PHYS-PUB-2016-005}  &  MG5\_\AMCatNLOV{2.3.3}~\cite{Alwall:2014hca} & NLO & \PYTHIAV{8.210-212}~\cite{Sjostrand:2014zea} & NLO~\cite{deFlorian:2016spz}  & NNPDF2.3LO~\cite{Ball:2012cx}  &  A14~\cite{ATL-PHYS-PUB-2014-021}  \\
 
$t\bar{t}WW$  &  MG5\_\AMCatNLOV{2.2.2}~\cite{Alwall:2014hca}  &  LO  &  \PYTHIAV{8.186}~\cite{Sjostrand:2007gs}  &  NLO~\cite{Alwall:2014hca}  & NNPDF2.3LO~\cite{Ball:2012cx}  & A14~\cite{ATL-PHYS-PUB-2014-021}  \\
 
$t\bar{t}WZ$  & MG5\_\AMCatNLOV{2.2.2}~\cite{Alwall:2014hca} & LO & \PYTHIAV{8.212}~\cite{Sjostrand:2014zea}  & NLO~\cite{Alwall:2014hca}  &  NNPDF2.3LO~\cite{Ball:2012cx}  &  A14~\cite{ATL-PHYS-PUB-2014-021}  \\
 
$tWZ$, $tZ$  & MG5\_\AMCatNLOV{2.3.3}~\cite{Alwall:2014hca} & LO & \PYTHIAV{8.212}~\cite{Sjostrand:2014zea}  & NLO~\cite{Alwall:2014hca}  &  NNPDF2.3LO~\cite{Ball:2012cx}  &  A14~\cite{ATL-PHYS-PUB-2014-021}  \\
 
$t\bar{t}H$~\cite{ATL-PHYS-PUB-2016-005}  &  \POWHEGV{2}~\cite{Nason:2004rx,Hartanto:2015uka}  &  NLO  &  \PYTHIAV{8.230}~\cite{Sjostrand:2014zea}  &  NLO~\cite{deFlorian:2016spz}  &  NNPDF2.3LO~\cite{Ball:2012cx}  &  A14~\cite{ATL-PHYS-PUB-2014-021} \\
 
$3t$, $4t$ 	&  MG5\_\AMCatNLOV{2.2.2}~\cite{Alwall:2014hca}  & LO &  \PYTHIAV{8.186}~\cite{Sjostrand:2007gs}  &  NLO~\cite{Alwall:2014hca}  &  NNPDF2.3LO~\cite{Ball:2012cx}  &  A14~\cite{ATL-PHYS-PUB-2014-021}  \\
 
\hline
\hline
 
$pp\to 4\ell,\,3\ell\nu$~\cite{ATL-PHYS-PUB-2017-005} & \SHERPAV{2.2.2}~\cite{Gleisberg:2008ta} &  NLO (0--1$j$)      & \SHERPA	 &  NLO  &  NNPDF3.0NNLO~\cite{Ball:2014uwa}  & \SHERPA \\
&                 &  + LO (2--3$j$)       &  &  &  & \\
$gg \to\ 4\ell$~\cite{ATL-PHYS-PUB-2017-005} & \SHERPAV{2.2.2}~\cite{Gleisberg:2008ta}      & LO (0--1$j$)  &  \SHERPA	 & NLO	&   NNPDF3.0 NNLO~\cite{Ball:2014uwa} & \SHERPA \\
$pp\to (2\ell2\nu/3\ell\nu/4\ell)jj$~\cite{ATL-PHYS-PUB-2017-005,ATL-PHYS-PUB-2019-004}  & \SHERPAV{2.2.2}~\cite{Gleisberg:2008ta}  &  LO  & \SHERPA		&      LO      & NNPDF3.0NNLO~\cite{Ball:2014uwa} & \SHERPA \\
 
$WH$, $ZH$	   & \PYTHIAV{8.186}~\cite{Sjostrand:2007gs}  &  LO  & \PYTHIA  &  LO  &  NNPDF2.3LO~\cite{Ball:2012cx}  &  A14~\cite{ATL-PHYS-PUB-2014-021}  \\
 
$VVV^{(*)}$  & \SHERPAV{2.2.1}~\cite{Gleisberg:2008ta}  & LO (0--1$j$)  & \SHERPA  &   LO     &     NNPDF3.0NNLO~\cite{Ball:2014uwa}	& \SHERPA \\
$VVVjj$~\cite{ATL-PHYS-PUB-2017-005} & \SHERPAV{2.2.2}~\cite{Gleisberg:2008ta}  & LO (0--1$j$)  & \SHERPA  &  LO &  NNPDF3.0NNLO~\cite{Ball:2014uwa}  &  \SHERPA \\
\hline\hline
\end{tabular}
}
\end{table}
 
The contributions of these processes to the SRs
are evaluated with Monte Carlo simulations
to determine the fiducial acceptance of the various regions
as well as the efficiencies of the detector and reconstruction software.
Table~\ref{tab:bkg-generators} provides a complete list
of the relevant processes considered in this analysis,
the event generators used for the predictions and their settings.
For the processes with the largest production cross-sections,
the scattering amplitudes evaluated for the event generation rely on terms up to the next-to-leading order (NLO) in the perturbative expansion,
while for other processes only leading-order (LO) terms are accounted for.
For most processes, the generated events are normalised to the inclusive cross-section computed with NLO accuracy,
either taken from the references indicated in Table~\ref{tab:bkg-generators}, or directly from the generator.
The generated events for the $t\bar tZ$, $tZ$, $tWZ$, $VZ$ and $VVZ$ processes include matrix elements for non-resonant $Z/\gamma^*\to\ell\ell$ contributions;
the same is true for non-resonant $W^*\to\ell\nu$ in events from $VV$ and $VVV$  processes.
For the \texttt{Rpc3LSS1b} SR,
only contributions from processes with three same-sign prompt leptons are evaluated with Monte Carlo simulations,
while the others ($VV$, $t\bar tV$\ldots) are included in the estimation of the reducible background, which is described in Section~\ref{sec:reducible_bkg}.
 
The generated events were processed through a detailed simulation of the ATLAS detector~\cite{SOFT-2010-01}
based on \textsc{Geant}4~\cite{Agostinelli:2002hh}.
To simulate the effects of additional $pp$ collisions in the same and nearby bunch crossings,
inelastic interactions were generated using the soft strong-interaction processes of \PYTHIAV{8.1.86}~\cite{Sjostrand:2014zea}
with a set of tuned parameters referred to as the A3 tune~\cite{ATL-PHYS-PUB-2016-017}
and the NNPDF23LO parton distribution function (PDF) set~\cite{Ball:2012cx}.
These inelastic interactions were overlaid onto the simulated hard-scatter events,
which were then reweighted to match the pile-up conditions observed in the data.
In all Monte Carlo samples, except those produced by the \SHERPA event generator,
the \textsc{EvtGen} 1.2.0 program~\cite{Lange:2001uf}
was used to model the properties of bottom and charm hadron decays.
 
Simulated events are weighted by scale factors
to correct for the mismodelling of inefficiencies associated with
the reconstruction of leptons, the application of lepton identification and isolation requirements,
the lepton-based trigger chains,
and the application of pile-up rejection (\texttt{JVT}) and $b$-tagging requirements
to jets that do or do not contain genuine $b$-flavoured hadrons.
 
Various sources of systematic uncertainties in the predicted event yields are accounted for.
Experimental sources, evaluated for all processes,
include uncertainties in the calibration of the momentum scale and resolution
for jets, leptons and the soft term of the missing transverse momentum,
as well as uncertainties in the various scale factors mentioned above,
in the measured integrated luminosity,
and in the distribution of the number of additional $pp$ interactions per event.
 
Uncertainties in the theoretical modelling of each process are also considered.
Uncertainties in the inclusive production cross-sections of $t\bar tW$, $t\bar tZ$ and $t\bar tH$
are taken as 12\%, 13\% and 8\%~\cite{deFlorian:2016spz}, respectively,
while a 6\% uncertainty is assigned for $VV$ processes~\cite{ATL-PHYS-PUB-2017-005}.
The impact of the choice of factorisation and renormalisation scales on the estimated fiducial acceptance and reconstruction efficiencies of the SRs
is assessed by considering the alternative event weights provided by the generators for up/down variations of these scales
(see e.g.\ Appendix B.3 in Ref.~\cite{Alwall:2014hca}).
The impact of PDF uncertainties is also taken into account by following the prescription in Ref.~\cite{Butterworth:2015oua}
using the sets of eigenvectors provided for each PDF~\cite{Ball:2012cx, Ball:2014uwa}.
 
For $t\bar tV$ and $t\bar tH$ processes,
the modelling of initial- and final-state radiation by the parton shower algorithm
is assessed by comparing five related variations of the \PYTHIAV{8} A14 event tune~\cite{ATL-PHYS-PUB-2014-021}.
For $t\bar tW$ the modelling of extra jets is further compared with
the prediction of the \SHERPAV{2.2.2} generator including LO matrix elements with two extra final-state partons;
the difference is found to be smaller than the tune-based parton shower uncertainties.
 
For $VV$ processes,
the impact of the choice of resummation scale (QSF) and CKKW matching scale~\cite{Hoeche:2009rj}
is also evaluated  by comparing the nominal prediction
with alternatives obtained with variations of these scales.
In addition, the modelling of high jet multiplicities is probed
by switching between different parton shower recoil schemes implemented in the \SHERPA generator~\cite{Hoeche:2009xc, Schumann:2007mg}.
 
Overall, modelling uncertainties in the SRs where these processes have sizeable contributions
are 35--45\% for $t\bar tW$, 25--45\% for $t\bar tZ$, 15--40\% for $t\bar tH$,
and 40--45\% for $WZ$.
For all other processes, uncertainties of $50\%$ are assigned.
The latter numbers are believed to be conservative
as these processes produce a larger number of jets at the first order of the perturbative expansion,
rendering them less sensitive to parton shower modelling uncertainties.
For the largest contribution to the SRs among these rarer processes, namely from $4t$ production,
the combined impact of factorisation and renormalisation scales as well as PDF uncertainties was checked and found to be indeed smaller than 50\%.
Modelling uncertainties are further assumed to be uncorrelated between processes shown in different categories in the tables and figures.
 
Three validation regions (VRs) enriched respectively in $WZ$+jets (\texttt{VRWZ4j}, \texttt{VRWZ5j})
and $t\bar tV$ (\texttt{VRttV})
are used to check the accuracy of the modelling of these processes
by comparing event yields predicted in a signal-free environment with data.
The definitions of these regions are provided in Table~\ref{tab:VR-definitions},
and are designed to minimise the level of reducible background.
Requirements are set on some of the variables defined in Section~\ref{sec:selection}.
The presence of a pair of same-flavour opposite-sign (SFOS) leptons is required in \texttt{VRWZ4j} and \texttt{VRWZ5j},
and its invariant mass $m_\text{SFOS}$ must be close to $m_Z$.
A minimum angular separation between the leading-\pt\ lepton and the jets ($\Delta R (\ell_1, j)$) is required in \texttt{VRttV},
together with a requirement on the ratio of the scalar $\pt$ sum over all $\btagged$ jets to the sum over all jets.
For all VRs, events belonging to any SR (except \texttt{Rpc3LSS1b}) are vetoed.
Upper bounds on \met\ and \meff\ are also imposed to minimise contributions from the benchmark SUSY scenarios of Figure~\ref{fig:benchmarks}.
Modelling uncertainties are evaluated with the same procedure as described above for the SRs,
and lead to uncertainties of around $20\%$ for $t\bar tV$ and $35\%$ for $WZ$ processes.
 
\begin{table}[t!]
\caption{
Event selection defining the three validation regions enriched in $WZ+$jets and $t\bar tV$ SM processes,
based on the variables defined in Section~\ref{sec:prompt_bkg}.
}
\label{tab:VR-definitions}
\def\arraystretch{1.1}
\resizebox{\textwidth}{!}{
\begin{tabular}{|c|c|c|c|c|c|}
\hline\hline
& $n_{\ell}$ & $n_b$  &  $n_j$ & \meff\ [GeV]  & Other requirements \\ \hline\hline
 
\texttt{VRWZ4j} & $=3$, & $=0$ & $\geq 4$ (\pt>\SI{25}{GeV}) & $>600$ &
$\SI{81}{GeV}<m_\text{SFOS}<\SI{101}{GeV}$, $\met>\SI{50}{GeV}$, \\ \cline{1-1}\cline{3-5}
\texttt{VRWZ5j} & $=1$ SFOS pair & $=0$ & $\geq 5$ ($\pt>\SI{25}{GeV}$) & $>400$ & no fourth baseline lepton \\
\hline\hline
& & & & & $\pt>\SI{30}{GeV}$ for the same-sign leptons, \\
\texttt{VRttV}  & $\geq 2$ ($\ell^\pm\ell^\pm$) & $\geq 1$ & $\geq 3$ ($\pt>\SI{40}{GeV}$) & $>600$ & $\sum{\pt^{b}} > 0.4 \sum{\pt^{j}}$, $\met > 0.1\meff$, \\
& & & & & $\Delta R_\eta (\ell_1, j) > 1.1$\\
\hline\hline
All VRs & \multicolumn{5}{|c|}{$\meff<\SI{1.5}{TeV}$, $\met<\SI{250}{GeV}$; veto \texttt{Rpc2L1b}, \texttt{Rpc2L2b}, \texttt{Rpc2L0b} and \texttt{Rpv2L} signal regions.}\\
\hline\hline
\end{tabular}
}
\end{table}
 
The number of events observed in each of the three VRs
and the corresponding predictions for SM processes
are shown in Table~\ref{tab:VR-yields},
including the reducible background described in the next section,
accounting for the systematic and statistical uncertainties.
The predicted event yields in all VRs are consistent with the data.
In the \texttt{VRWZ4j} and \texttt{VRWZ5j} regions,
the large systematic uncertainties include contributions from theoretical modelling
and from experimental sources (dominated by the jet energy scale) due to the large required number of jets.
 
 
\begin{table}[t!]
\caption{
Observed yields in data compared with the expected contributions from relevant SM processes (section~\ref{sec:prompt_bkg}) and the reducible background (section~\ref{sec:reducible_bkg}),
in the three VRs enriched in $WZ+$jets and $t\bar tV$ processes.
The displayed numbers include all sources of statistical and systematic uncertainties;
since some of the latter might be correlated between different processes,
the numbers do not necessarily add up in quadrature to the uncertainty in the total expected background.
Selections with three leptons are not affected by the charge-flip electron background,
so such contributions are denoted by $-$.
}
\label{tab:VR-yields}
\begin{center}
\setlength{\tabcolsep}{0.0pc}{
\small
\begin{tabular*}{\textwidth}{@{\extracolsep{\fill}}lccc}
\noalign{\smallskip}\hline\hline\noalign{\smallskip}
 
&    \texttt{VRttV}    &    \texttt{VRWZ4j}    &    \texttt{VRWZ5j}    \\[-0.05cm]
 
\noalign{\smallskip}\hline\noalign{\smallskip}
 
Observed    &    127    &    355    &    190    \\
 
\noalign{\smallskip}\hline\hline\noalign{\smallskip}
 
Total SM background    &    \hphantom{.}106$^{+16\hphantom{0.}}_{-19}$
&    \hphantom{.}390$^{+120\hphantom{.}}_{-100}$
&    \hphantom{.}209$^{+68\hphantom{0.}}_{-54}$  \\
 
\noalign{\smallskip}\hline\noalign{\smallskip}
 
$t\bar{t}W$     &    25.8$^{+5.5\hphantom{0}}_{-5.6}$
&    0.40$^{+0.17}_{-0.15}$
&  0.32$^{+0.14}_{-0.15}$    \\[5pt]
 
$t\bar{t}Z$     &    34.4$^{+8.1\hphantom{0}}_{-8.2}$
&    37.2$^{+8.6\hphantom{0}}_{-8.8}$
&    27.3$^{+7.2\hphantom{0}}_{-7.4}$     \\[5pt]
 
$WZ$     &    \hphantom{0}5.8$^{+2.5\hphantom{0}}_{-2.2}$
&    \hphantom{.}310$^{+120\hphantom{.}}_{-90}$
&    \hphantom{.}153$^{+64\hphantom{0.}}_{-50}$    \\[5pt]
 
$ZZ$, $W^\pm W^\pm$, $VH$, $VVV$     &    1.03$^{+0.40}_{-0.39}$
&    12.0$^{+3.4\hphantom{0}}_{-2.9}$
&    \hphantom{0}7.5$^{+2.8\hphantom{0}}_{-2.1}$     \\[5pt]
 
$t\bar tH$     &    \hphantom{0}7.3$^{+1.1\hphantom{0}}_{-1.2}$
&    0.90$^{+0.18}_{-0.18}$
&    0.81$^{+0.18}_{-0.17}$   \\[5pt]
 
$t(W)Z$, $t\bar tVV$, $3t$, $4t$     &    10.4$^{+5.2\hphantom{0}}_{-5.2}$
&    10.3$^{+5.3\hphantom{0}}_{-5.3}$
&    \hphantom{0}5.8$^{+3.1\hphantom{0}}_{-3.1}$   \\[5pt]
 
Fake/non-prompt     &    \hphantom{0.}14$^{+8\hphantom{00.}}_{-12}$
&    \hphantom{0.}15$^{+7\hphantom{00.}}_{-13}$
&    13.7$^{+5.4\hphantom{0}}_{-8.0}$    \\[5pt]
 
Charge-flip     &    \hphantom{0}7.1$^{+5.7\hphantom{0}}_{-5.7}$
&    $-$    &    $-$    \\[5pt]
 
\noalign{\smallskip}\hline\hline\noalign{\smallskip}
\end{tabular*}
}
\end{center}
\end{table}
 
Other potential sources of same-sign leptons in the SRs are not included,
as they were estimated to be negligible.
These include simultaneous production of massive bosons or top quarks
via either double parton scattering (DPS) or pile-up interactions.
Simple estimations of the inclusive production cross-sections were performed for several processes.
For DPS the approach from Ref.~\cite{Humpert:1984ay} was used, relying on the DPS effective cross-section $\sigma_\text{eff}$~\cite{Treleani:2007gi}.
Earlier experiments probed the reliability of this approach for different centre-of-mass energies and physics processes~\cite{STDM-2017-15},
including more recently for $W^\pm W^\pm$ production~\cite{CMS-SMP-18-015}.
All these measurements display a level of consistency allowing to conclude that DPS processes would not contribute noticeably to the SRs.
For pile-up interactions,
the estimation was based on the longitudinal density of reconstructed vertices~\cite{ATL-PHYS-PUB-2018-007},
as the impact parameter requirements in the selection of the leptons strongly affect the yields of such processes.
The only process for which the pile-up induced contribution
is estimated to be more than $1\%$ of the corresponding SM process
is $W^\pm W^\pm$ production,
which has been highlighted~\cite{Gaunt:2010pi} as a sensitive process for DPS measurements.
But this process is in itself a minor source of background for this analysis.
 
Another source, notably highlighted in Ref.~\cite{Onyisi:2017mza},
is the production of additional pairs of leptons in radiative top quark decays,
$t\to \ell\nu b\ell\ell$ or $t\to qq b\ell\ell$,
which are not included in the generator matrix elements for the $t\bar tZ$ process.
These contributions were studied by running the \textsc{Photos}++ QED shower program~\cite{Davidson:2010ew}
on the tree-level decay products of top quarks generated with \MADGRAPHV{2.6} or \PYTHIAV{8}.
The fraction of events in which an additional lepton is produced drops sharply with the \pt\ requirement for that lepton;
for a $\pt=\SI{10}{GeV}$ threshold\footnote{
Dilepton $t\bar t$ events with an extra $\pt>\SI{10}{GeV}$ lepton
satisfy the lepton selection requirements of this analysis.
}
this fraction was found to be $\sim 0.2\%$, a similar order of magnitude to that quoted in Ref.~\cite{Davidson:2010ew}.
An additional isolation requirement similar to that used in the analysis reduces this rate by a factor of three.
This represents less than $2\%$ ($4\%$) of the inclusive contribution from $t\bar tV$ processes
for final states with same-sign (three) leptons;
furthermore, the smaller reconstruction and identification efficiencies for low-\pt\ leptons
should further reduce the radiative top quark decay contribution relative to $t\bar{t}V$ processes.
The expected contribution to the SRs is therefore small enough to be neglected.
 
\section{Backgrounds with non-prompt, fake or charge-flip leptons}
\label{sec:reducible_bkg}
Other SM processes that do not lead to genuine production of same-sign prompt leptons,
such as \ttbar\ processes and to a much lesser extent production of $W/Z+$jets or single top quarks,
might contaminate the SRs via secondary interactions,
for example bremsstrahlung or non-prompt leptons in ensuing decays,
or misidentification of the reconstructed objects (fakes).
 
The first source consists of `charge-flip' electrons,
where the charge of a prompt electron is mismeasured
due to the emission of a bremsstrahlung photon which through interaction with detector material converts into a pair of secondary electron tracks,
one of which happens to better match the position of the calorimeter cluster than the original electron track
and has a charge opposite to that of the prompt electron.
Thanks to the application of the \texttt{ECIDS} discriminant for signal electrons,
charge-flip electrons are only a minor background in the SRs.
Muon charge-flip is negligible in the \pt\ range relevant to this analysis.
 
Backgrounds with charge-flip electrons are estimated by selecting data events with two opposite-sign leptons,
and weighting them by the probability of one electron charge to be mismeasured.
This offers a large improvement in statistical accuracy over relying directly on the simulation for these backgrounds,
as well as the elimination of associated experimental and theory uncertainties.
The charge-flip probabilities are measured in simulated $t\bar t$ events,
as a function of \pt\ and $\abseta$.
They are corrected by scale factors corresponding to the ratio of probabilities
measured in data and simulation
from the reconstructed charges of electrons produced in $Z\to e^+e^-$ decays and selected with a `tag and probe' method~\cite{EGAM-2018-01}.
The probabilities reach $\mathcal{O}(0.1\%)$ at $\pt=\SI{100}{GeV}$ for central electrons ($\abseta<1.4$),
and are about fives times larger at higher $\abseta$
due to the larger amount of material traversed by electrons.
Systematic uncertainties are assessed by propagating the measurement uncertainties,
leading to a 70--90\% uncertainty in the predicted SR yields for the charge-flip background.
 
The data weighting method described above neglects the differences in momentum scale and resolution between standard and charge-flip electrons.
This approximation was validated by recomputing the expected SR yields
after reducing the \pt\ of the electron with largest $\abseta$ by \SI{5}{GeV} ---
a value bounding from above the invariant mass resolution of same-sign $ee$ pairs near the Z boson mass ---
in all weighted data events,
which was found to have a negligible impact on the results.
For the \texttt{Rpc3LSS1b} SR,
the method is adapted by simply selecting data events with three or more leptons,
which are weighted by the probability of one or more electron charges to be mismeasured
such that the resulting event contains three same-sign leptons.
 
Another, more important, source of reducible background includes fake or non-prompt leptons,
referred to in the following as `F/NP' leptons.
These may originate from electroweak-mediated decays of hadrons (in particular $b$- and $c$-flavoured hadrons in decays of top quarks and weak bosons),
single pions stopped in the EM calorimeter that fake electron signatures,
in-flight decays of kaons into muons,
or the conversion of photons into pairs of electrons in the beam pipe or detector material.
Lepton candidates reconstructed from these different sources share the properties of being generally not well isolated
and being mostly rejected by the lepton identification criteria and impact parameter requirements.
Therefore, all sources of background with F/NP leptons are estimated together,
using a common method that exploits these properties.
 
Sources of F/NP leptons in the SRs are mostly semileptonic or dileptonic $t\bar t$ processes.
To estimate their contributions to the SRs,
a matrix method as described in Ref.~\cite{SUSY-2013-09} is used,
with a different parameterisation of efficiencies and uncertainties as detailed in the following.
It relies on data events selected with the same criteria as in the region of interest,
but with a loosened lepton selection corresponding to the baseline leptons defined in Section~\ref{sec:reconstruction}
after the overlap removal procedure with a few extra adjustments:
muons are required to satisfy a loosened transverse impact parameter requirement $|d_0|<7\sigma(d_0)$,
and electrons must both be within $\abseta<2.0$ and satisfy the \texttt{ECIDS} requirement against charge-flip.
These adjustments align the selection with the fiducial acceptance of signal leptons,
and eliminate irrelevant sources of reconstructed leptons.
The matrix method, for the simplest situation where selected events contain a single lepton,
relies on the following asymptotic equality for
the observed proportion of events $\mathcal S$ where the lepton satisfies the signal lepton requirements:
 
\begin{equation}
\mathcal{S} = \varepsilon\left(1-\mathcal{F}\right) + \zeta \mathcal{F}
\label{eq:matrix}
\end{equation}
 
where $\mathcal{F}$ is the unknown proportion of events with a F/NP lepton,
while $\varepsilon$ and $\zeta$ are respectively the probabilities for a prompt or F/NP lepton to satisfy the signal lepton requirements.
If $\varepsilon$ and $\zeta$ are known,
Eq.~(\ref{eq:matrix}) can be used to determine $\mathcal{F}$ and thus the number of events with a F/NP lepton in the region of interest.
The approach can be generalised to events with arbitrary numbers of leptons,
as well as the more realistic situation where $\varepsilon$ and $\zeta$ depend on the flavour and kinematic properties of the leptons.
 
The probabilities $\varepsilon$ are obtained directly from the $t\bar t$ simulation,
as a function of \pt\ and $\abseta$,
accounting for the various lepton-related scale factors mentioned in Section~\ref{sec:prompt_bkg}.
For $\pt>\SI{30}{GeV}$ the probabilities are larger than $80\%$ and $90\%$ for electrons and muons, respectively.
As $\varepsilon$ might be smaller in data events coming from signal scenarios
with busy environments, such as boosted top quarks that decay semileptonically,
uncertainties are taken into account as a function of \pt\ and the proximity to the closest jet
and can be as large as $30\%$ for $\Delta R<0.4$.
 
The probabilities $\zeta$ are measured in regions of the data
enriched in F/NP leptons produced by $t\bar t$ processes,
defined by selecting events with two same-sign leptons or three leptons, at least one \btagged jet,
$\met>\SI{30}{GeV}$ and $\geq$ 2--3 jets;
upper bounds on \met\ and \meff\ avoid contamination from supersymmetric processes.
The probabilities are measured as a function of \pt,
separately for events with exactly one or exactly two \btagged jets,
as the proportion of non-prompt leptons from $b$-flavoured hadron decays is much smaller in the latter case
than in events with at most one \btagged jet.
They are also measured separately for electrons that were or were not used to accept the event via a lepton-based trigger,
as the requirements for electrons reconstructed online differ from those for electrons reconstructed offline.
The measured probabilities are $\sim 10\%$ for both electrons and muons up to $\pt\sim\SI{35}{GeV}$,
and increase to $20\%$ and $35\%$ for electrons and muons with $\pt>\SI{60}{GeV}$.
They can be up to twice as large in events with two \btagged jets.
 
Systematic uncertainties in the measured $\zeta$ probabilities
account for variations in the relative contributions of different sources of F/NP leptons or in the environment,
and they are assessed in simulated $t\bar t$ events.
For electrons the latter amount to a $30\%$ additional uncertainty.
For muons the probabilities become smaller in events with a larger amount of activity,
where non-prompt muons tend to be less well isolated.
This leads to extra uncertainties of $30\%$ to $80\%$ for $\pt>\SI{50}{GeV}$,
as this effect is not accounted for with the simple \pt-based parameterisation used for $\zeta$.
 
Events with charge-flip electrons may bias the matrix method prediction,
as the probability for such electrons to satisfy signal lepton requirements
differ from both standard and F/NP electrons.
For that reason, estimated charge-flip contributions are subtracted from the data event yields when the method is applied.

\begin{figure}[t!]
\begin{center}
\includegraphics[width=\textwidth]{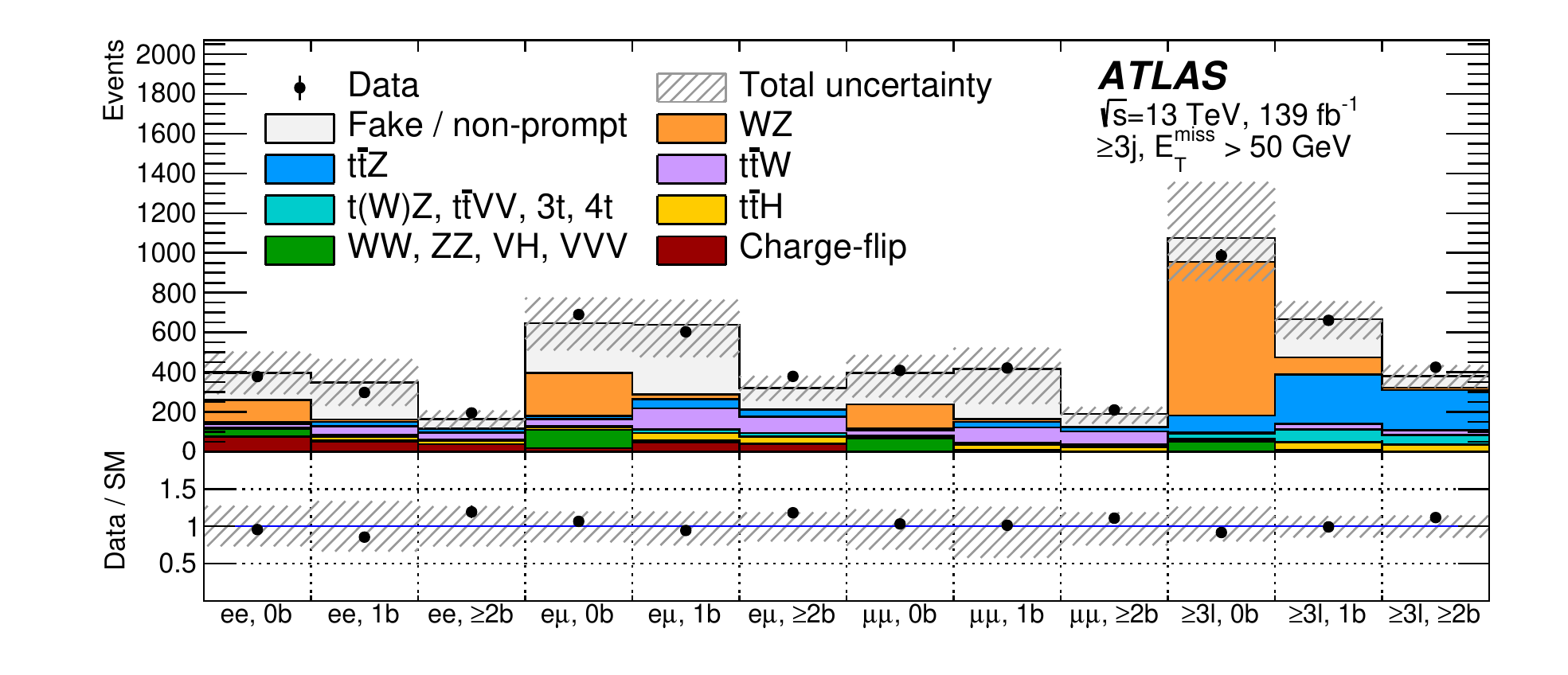}
\end{center}
\caption{
Data event yields compared with the expected contributions from
relevant SM processes (section~\ref{sec:prompt_bkg}) and the reducible background (section~\ref{sec:reducible_bkg}),
after a loose preselection
requiring events with same-sign leptons, $\met>\SI{50}{GeV}$ and at least three jets with $\pt>\SI{40}{GeV}$.
The observed and predicted event yields are classified as a function of the number and flavour of the leptons,
as well as the number of \btagged jets.
The uncertainties, shown with hashed bands,
include the total uncertainties in the reducible background,
as well as the modelling and statistical uncertainties for the Monte Carlo simulations.
}
\label{fig:preselection-yields}
\end{figure}

\begin{figure}[t!]
\begin{center}
\includegraphics[width=0.49\textwidth]{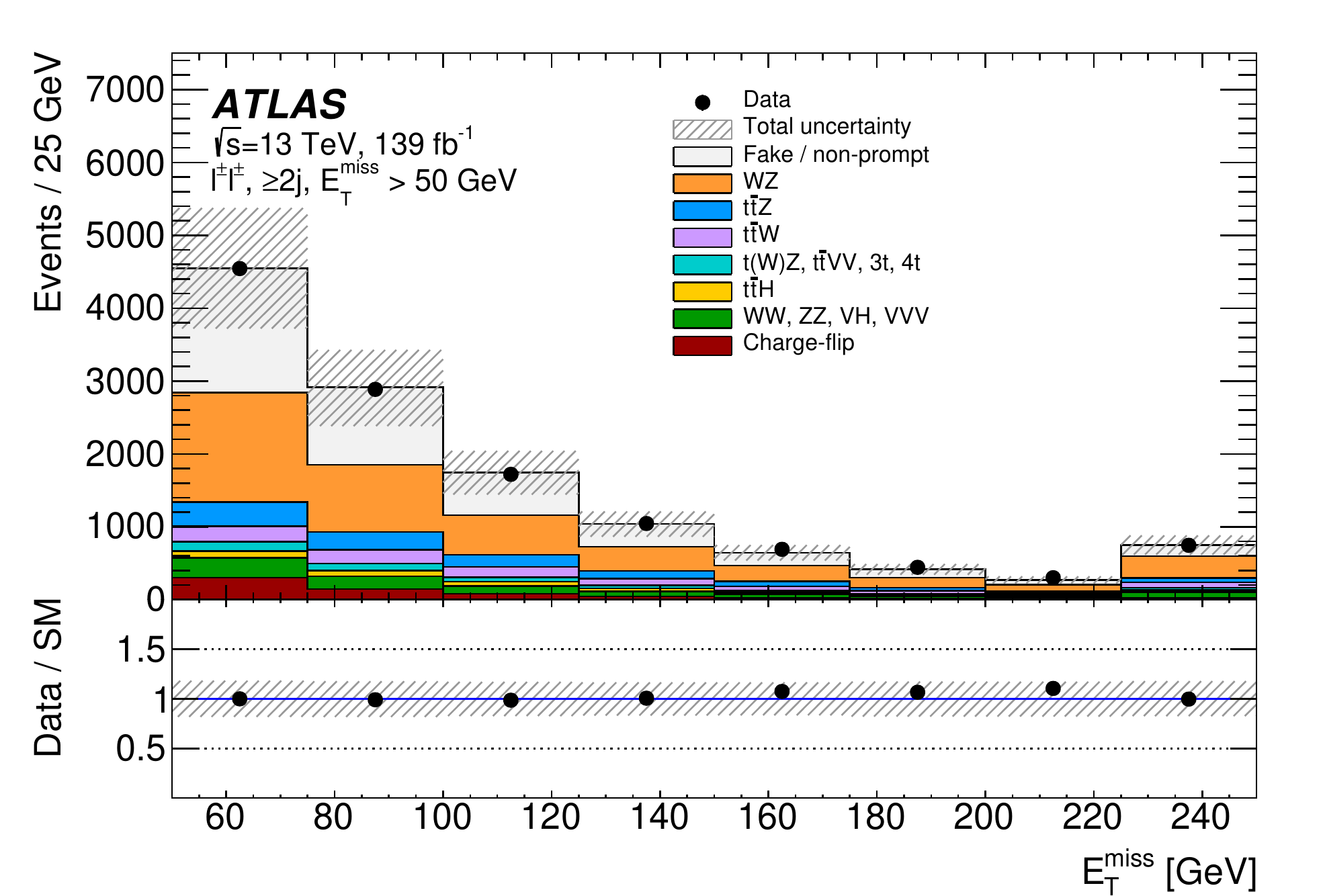}
\includegraphics[width=0.49\textwidth]{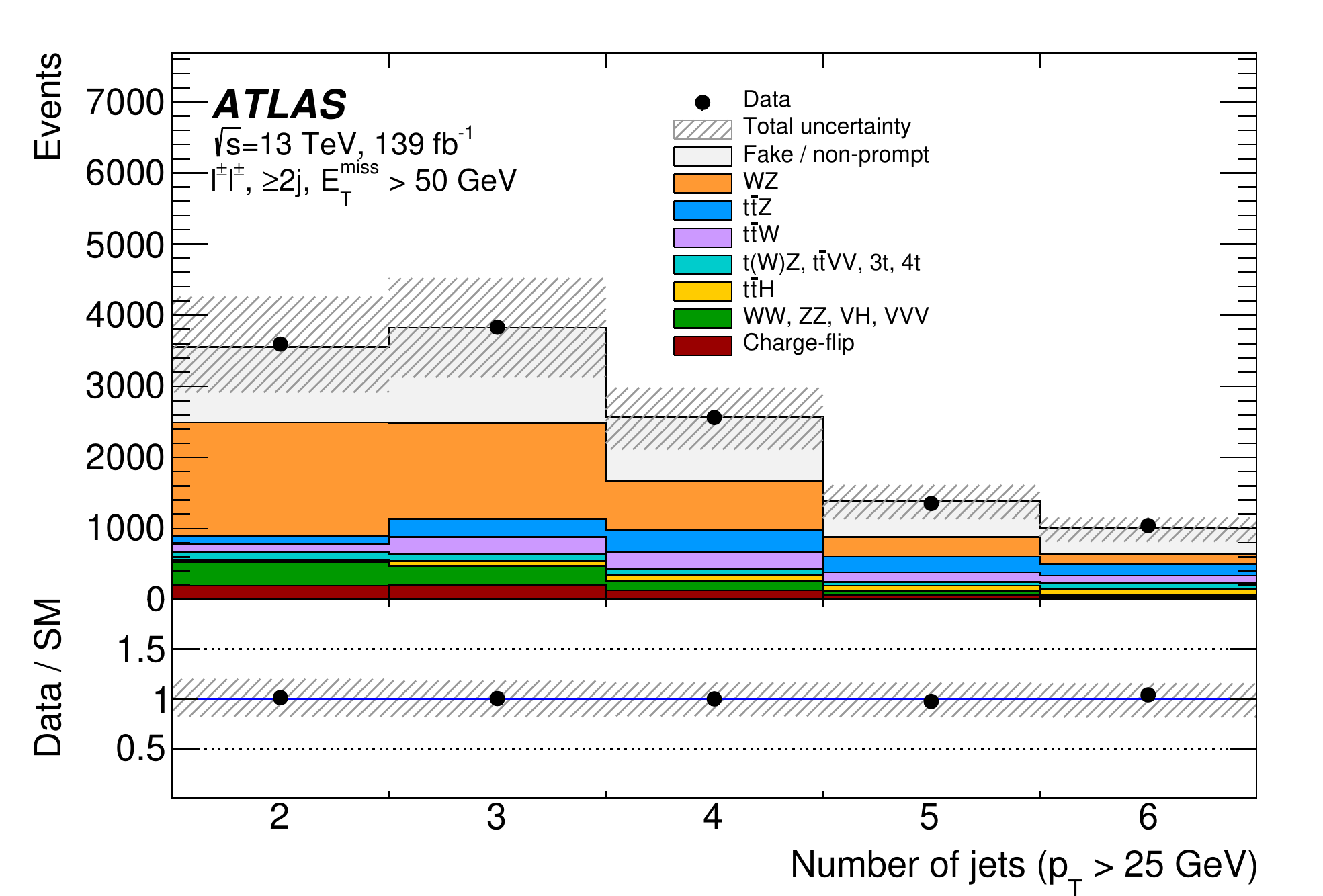}
\end{center}
\caption{
Distributions of (left) \met\ and (right) the number of jets with $\pt>\SI{25}{GeV}$,
after a loose preselection
requiring events with same-sign leptons, $\met>\SI{50}{GeV}$ and at least two jets with $\pt>\SI{40}{GeV}$.
The uncertainties, shown with hashed bands,
include the total uncertainties in the reducible background,
as well as the modelling and statistical uncertainty for the Monte Carlo simulations.
The last bin is inclusive.
}
\label{fig:distributions}
\end{figure}

The data-driven methods employed to estimate the reducible background
are validated by comparing the event yields in data with the combined predictions for these backgrounds,
added to Monte Carlo predictions for SM processes as described in Section~\ref{sec:prompt_bkg}.
Figure~\ref{fig:preselection-yields} shows such a comparison
for a loose event preselection requiring
same-sign leptons, $\met>\SI{50}{GeV}$ and at least three jets with $\pt>\SI{40}{GeV}$,
binned in the different lepton flavour and $b$-tag multiplicity combinations.
Simulation studies show that the sources of reducible background
for such a preselection are dominated, as in the SRs, by $t\bar t$ processes.
While the F/NP lepton background represents a major contribution to the total yields, the charge-flip background is always small.
In all bins, the observed and predicted event yields agree within uncertainties.
Figure~\ref{fig:distributions} presents the distributions of \met\ and the number of jets
in events with at least two jets and an otherwise identical preselection,
for which good agreement is observed between data and predictions.
 
As $t\bar t$ processes with F/NP leptons produce a major background in this analysis,
the estimated SR yields obtained with the matrix method are cross-checked against an alternative method based on a factorisation approach.
In the latter, a control region CR is built for each SR by relaxing some of the \met\ or \meff\ requirements.
Another set of regions SR$'$ and CR$'$ is built with identical criteria
but using events where a single lepton is selected instead of the same-sign pair,
as well as an additional object (jet, $b$-tagged jet, photon) that might be a source of F/NP leptons.
Each CR is chosen such that the kinematic properties of the additional object are similar in CR$'$ and SR$'$.
A `transfer factor' is built as the ratio of the number of data events in SR$'$ to the number in CR$'$.
The transfer factor is then multiplied by the number of events with same-sign leptons in the CR
to obtain an estimate of the F/NP lepton background yield in the SR.
The estimated contributions to the CR from SM processes with same-sign prompt leptons are subtracted, as is the charge-flip background.
Differences between the transfer factors calculated using different choices for the additional object are
treated as a source of systematic uncertainty.
The estimated F/NP lepton background yields in the five SRs obtained with this alternative method
are consistent with the matrix method prediction within uncertainties.
 
\section{Results}
\label{sec:results}
\begin{figure}[!hpt]
\includegraphics[width=\textwidth]{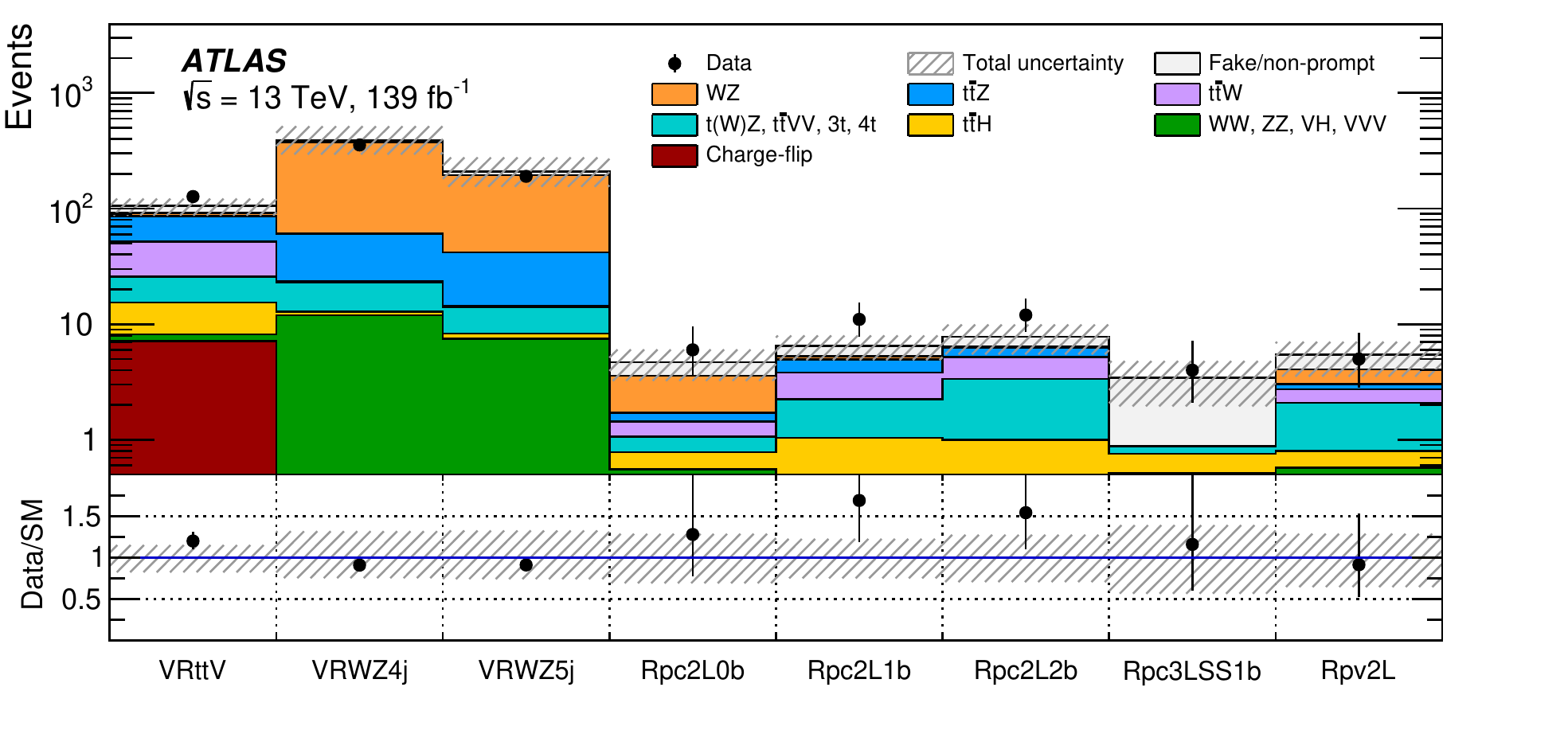}
\caption{
Data event yields compared with the expected contributions from
relevant SM processes (section~\ref{sec:prompt_bkg}) and the reducible background (section~\ref{sec:reducible_bkg}),
in the three VRs and the five SRs.
The total uncertainties in the expected event yields are shown as the hashed bands.
}
\label{fig:SR-VR-yields}
\end{figure}
 
 
\begin{table}[!hpb]
\caption{
Observed yields in data and expected contributions from SM processes (section~\ref{sec:prompt_bkg}) and the reducible background (section~\ref{sec:reducible_bkg}) to the five SRs.
The displayed numbers include all sources of statistical and systematic uncertainty;
since some of the latter might be correlated between different processes,
the numbers do not necessarily add up in quadrature to the uncertainty in the total expected background.
The $WZ$ and $t\bar tV$ processes cannot genuinely result in final states with three same-sign leptons,
so their contributions to the \texttt{Rpc3LSS1b} signal region are denoted by $-$.
Contributions to \texttt{Rpc3LSS1b} only include those from processes
producing final states with three genuine same-sign leptons, such as $t\bar tWZ$ or $WZZ$.
}
\label{tab:SR-yields}
 
\setlength{\tabcolsep}{0.0pc}{
\small
\begin{tabular*}{\textwidth}{@{\extracolsep{\fill}}lccccc}
\noalign{\smallskip}\hline\hline\noalign{\smallskip}
 
&    \texttt{Rpc2L0b}    &    \texttt{Rpc2L1b}    &    \texttt{Rpc2L2b}    &    \texttt{Rpc3LSS1b}    &    \texttt{Rpv2L}    \\[-0.05cm]
 
\noalign{\smallskip}\hline\noalign{\smallskip}
 
Observed    &    6    &    11    &    12    &    4    &    5    \\
 
\noalign{\smallskip}\hline\hline\noalign{\smallskip}
 
Total SM background    &    \hphantom{0}4.7$^{+1.3\hphantom{0}}_{-1.5}$
&    \hphantom{0}6.5$^{+1.5\hphantom{0}}_{-1.6}$
&    \hphantom{0}7.8$^{+2.1\hphantom{0}}_{-2.3}$
&    \hphantom{0}3.5$^{+1.4\hphantom{0}}_{-1.5}$
&    \hphantom{0}5.5$^{+1.6\hphantom{0}}_{-2.0}$ \\[5pt]
 
\noalign{\smallskip}\hline\noalign{\smallskip}
 
$t\bar{t}W$     &    0.38$^{+0.21}_{-0.22}$
&    1.56$^{+0.64}_{-0.63}$
&    1.81$^{+0.74}_{-0.62}$
&    $-$
&    0.64$^{+0.30}_{-0.29}$ \\[5pt]
 
$t\bar{t}Z$     &    0.26$^{+0.13}_{-0.11}$
&    1.17$^{+0.43}_{-0.43}$
&    1.04$^{+0.34}_{-0.33}$
&    $-$
&    0.30$^{+0.18}_{-0.17}$ \\[5pt]
 
$WZ$     &    1.88$^{+0.85}_{-0.76}$
&    0.29$^{+0.15}_{-0.15}$
&    0.21$^{+0.10}_{-0.11}$
&    $-$
&    1.03$^{+0.48}_{-0.45}$ \\[5pt]
 
$ZZ$, $W^\pm W^\pm$, $VH$, $VVV$ &   0.50$^{+0.18}_{-0.16}$
&    0.04$^{+0.02}_{-0.02}$
&    0.03$^{+0.01}_{-0.01}$
&    $<0.02$
&    0.43$^{+0.13}_{-0.13}$ \\[5pt]
 
$t\bar tH$  & 0.23$^{+0.09}_{-0.08}$
& 0.90$^{+0.27}_{-0.26}$
& 0.75$^{+0.25}_{-0.20}$
& 0.24$^{+0.05}_{-0.05}$
& 0.22$^{+0.10}_{-0.09}$ \\[5pt]
 
$t(W)Z$, $t\bar tVV$, $3t$, $4t$  &  0.29$^{+0.17}_{-0.16}$
&    1.21$^{+0.63}_{-0.63}$
&    \hphantom{0}2.4$^{+1.2\hphantom{0}}_{-1.2}$
&    0.12$^{+0.06}_{-0.06}$
&    1.29$^{+0.67}_{-0.67}$ \\[5pt]
 
Fake/non-prompt     &    \hphantom{0}1.1$^{+0.8\hphantom{0}}_{-1.1}$
&    \hphantom{0}1.3$^{+0.9\hphantom{0}}_{-1.1}$
&    \hphantom{0}1.4$^{+1.4\hphantom{0}}_{-1.4}$
&    \hphantom{0}2.6$^{+1.3\hphantom{0}}_{-1.5}$
&    \hphantom{0}1.4$^{+1.2\hphantom{0}}_{-1.4}$ \\[5pt]
 
Charge-flip     &    0.05$^{+0.04}_{-0.04}$
&    0.11$^{+0.11}_{-0.11}$
&    0.22$^{+0.22}_{-0.22}$
&    0.52$^{+0.39}_{-0.39}$
&    0.14$^{+0.14}_{-0.14}$ \\[5pt]
\noalign{\smallskip}\hline\hline\noalign{\smallskip}
\end{tabular*}
}
\end{table}
 
\begin{figure}[!t]
\begin{center}
\includegraphics[width=0.49\textwidth]{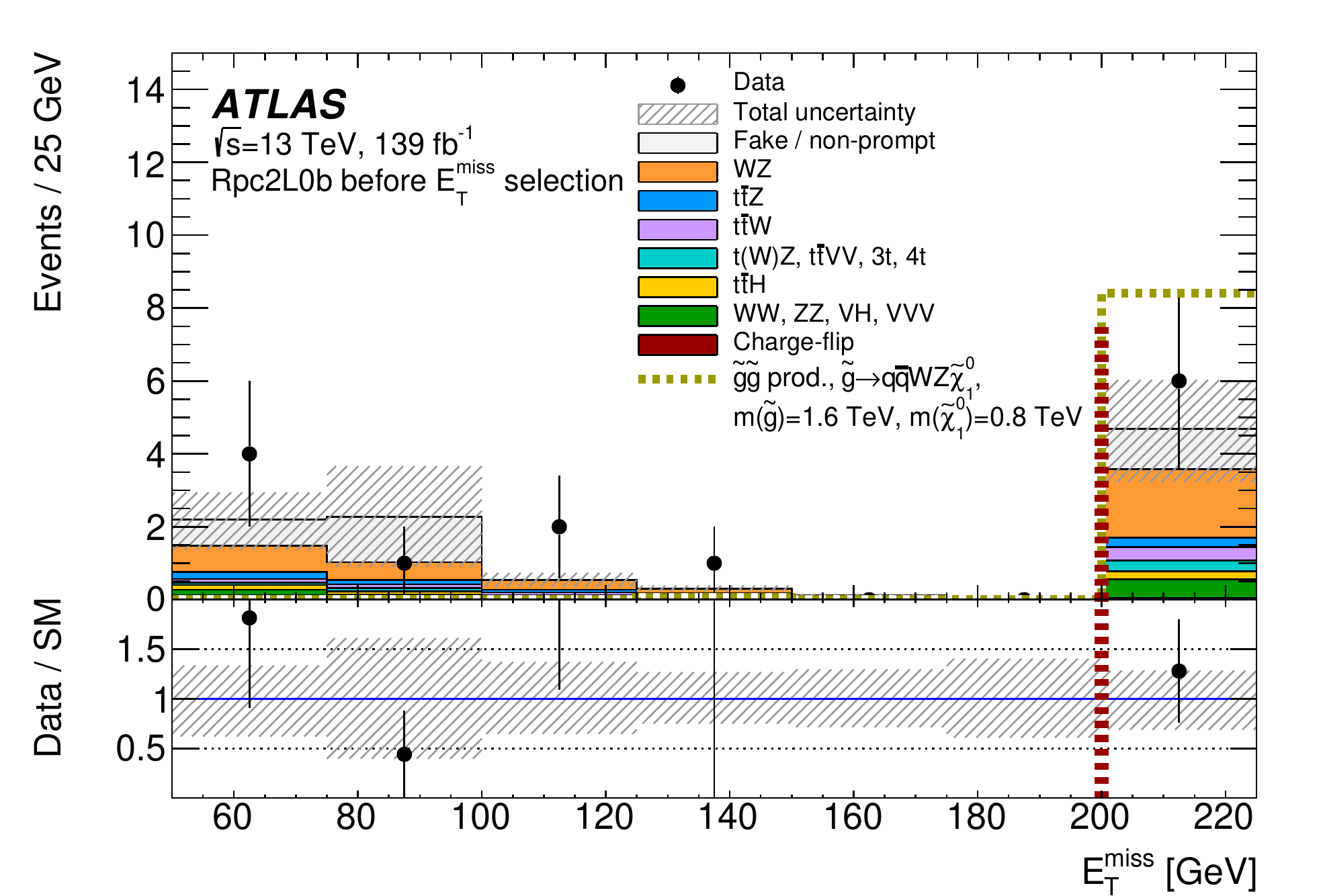}
\includegraphics[width=0.49\textwidth]{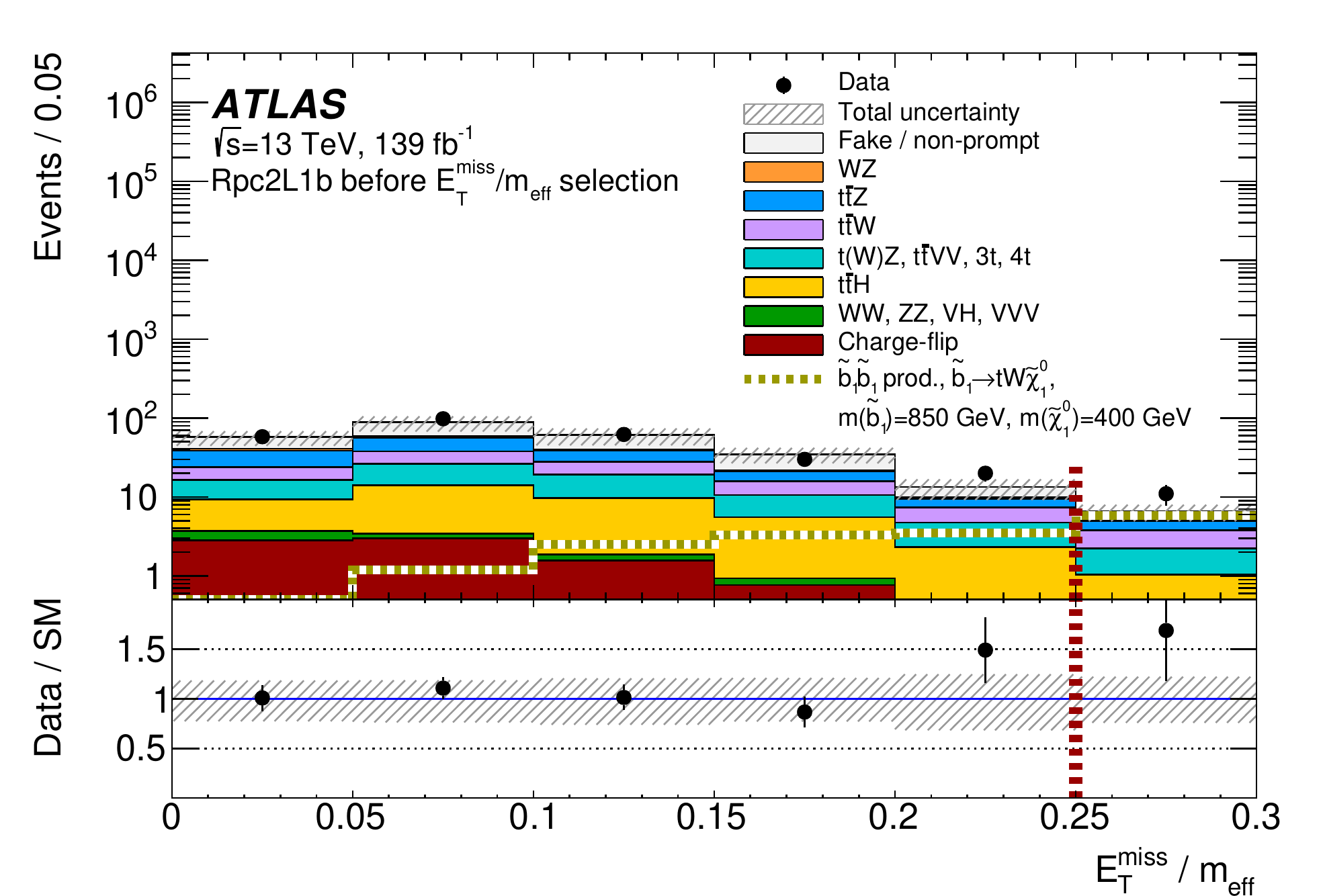}
 
\includegraphics[width=0.49\textwidth]{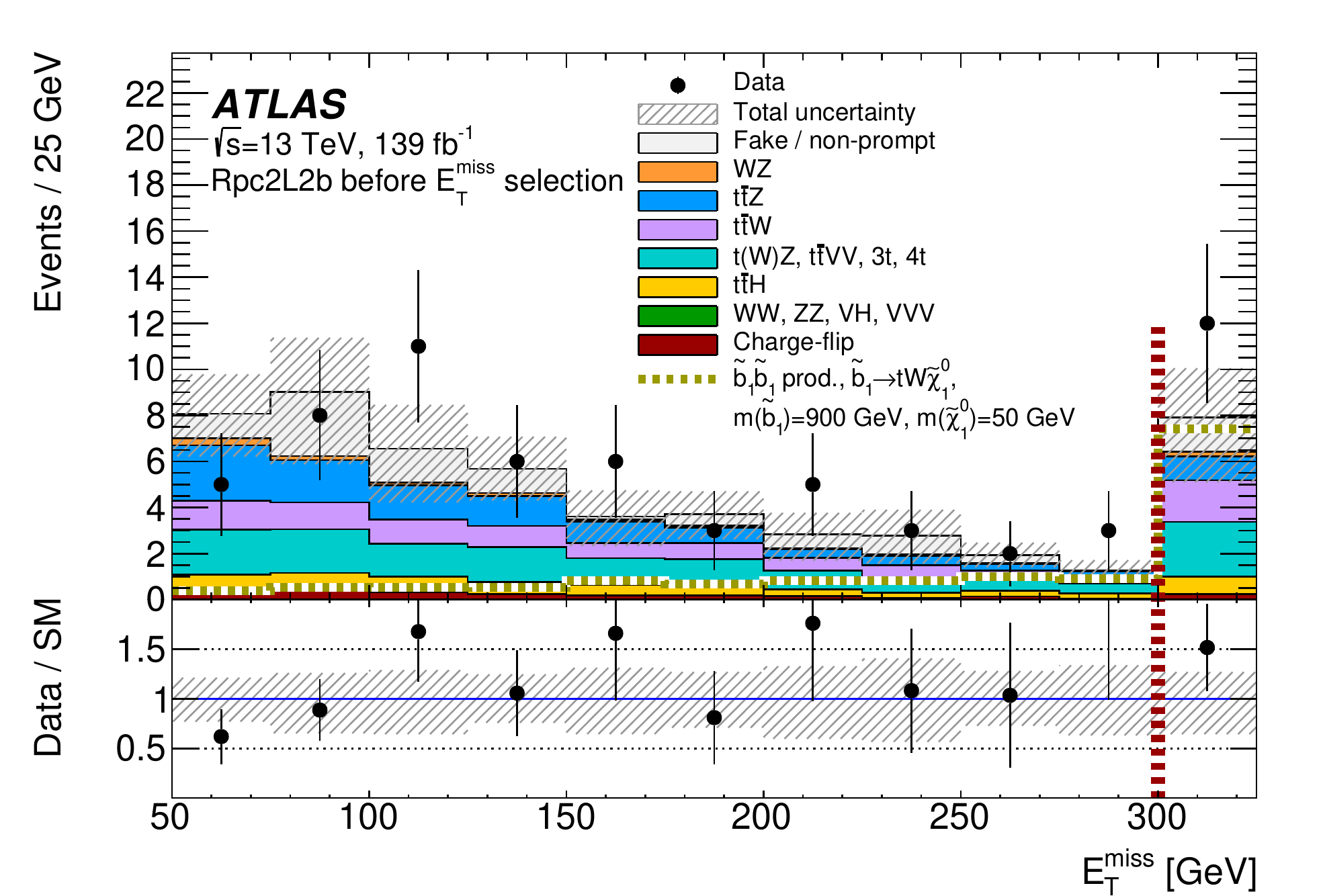}
\includegraphics[width=0.49\textwidth]{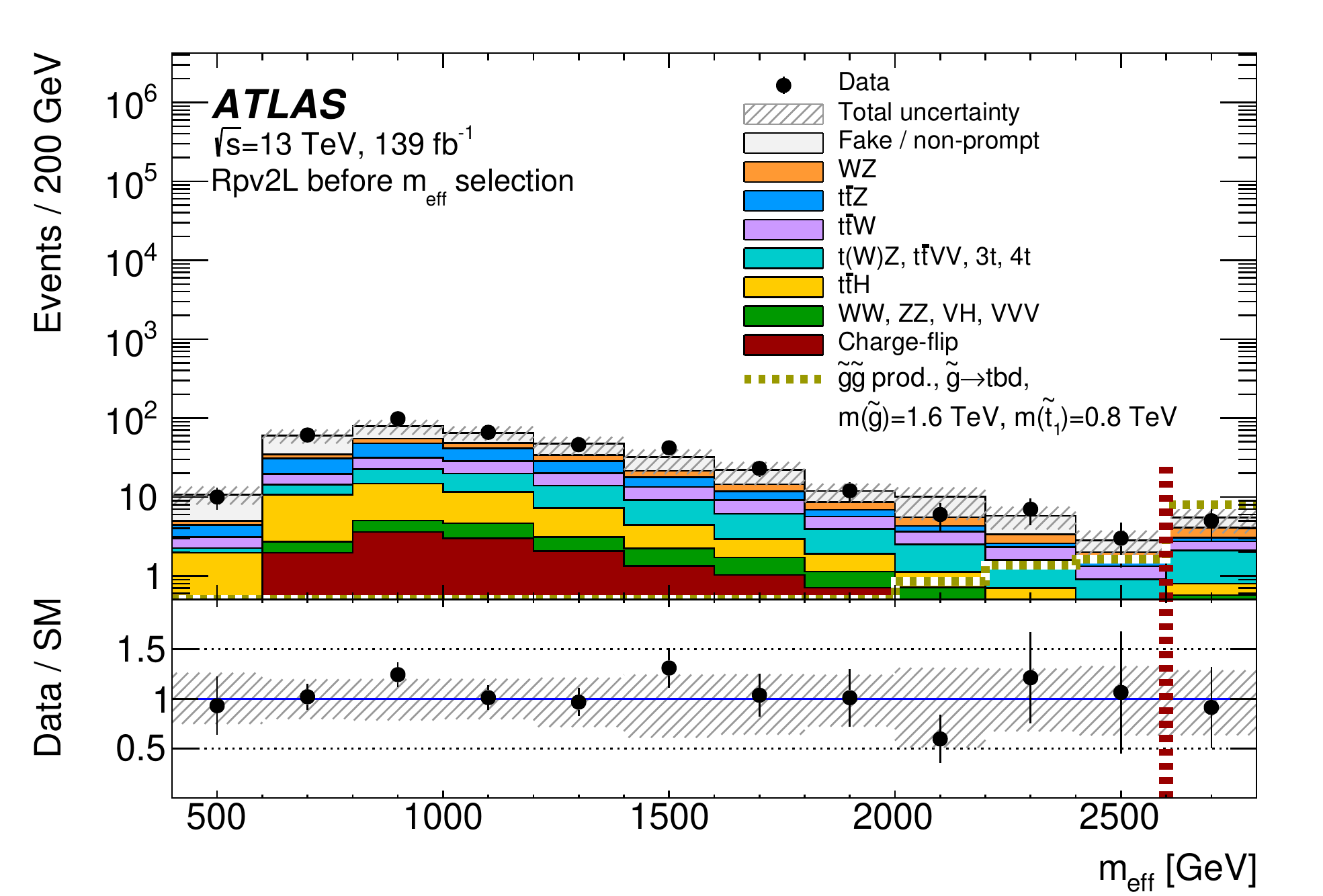}
\end{center}
\caption{
Distributions of \met, \meff or the $\met/\meff$ ratio
near the SRs (top left) \texttt{Rpc2L0b},
(top right) \texttt{Rpc2L1b},
(bottom left) \texttt{Rpc2L2b}
and (bottom right) \texttt{Rpv2L}.
The total uncertainties in the expected event yields are shown as the hashed bands.
The last bin, isolated by a vertical red dashed line, is inclusive and corresponds to the SR.
Hypothetical contributions from representative SUSY scenarios are displayed by the dashed-line overlaid histograms.
}
\label{fig:SR-nminus1}
\end{figure}

 
The event yields in data in the five SRs,
and the corresponding estimates for SM processes and the reducible background,
are shown in Figure~\ref{fig:SR-VR-yields} and detailed in Table~\ref{tab:SR-yields}.
No significant excess over the expected yields is observed in any of the SRs.
The SRs \texttt{Rpc2L1b} and \texttt{Rpc2L2b} overlap by approximately $15\%$
in terms of expected yields from SM processes,
and two data events satisfy the requirements for both regions.
Among SM processes with smaller cross-sections,
the largest contributions originate from $t\bar{t}H$
(in \texttt{Rpc2L1b}) and $4t$ (in \texttt{Rpc2L2b}, \texttt{Rpv2L}).
The distributions of \met, \meff or the $\met/\meff$ ratio
are shown
with the SR requirement relaxed for the displayed variable
in Figure~\ref{fig:SR-nminus1} for four of the SRs.
When \met\ is relaxed (\texttt{Rpc2L0b}, \texttt{Rpc2L2b}),
the $\meff$ requirement is also loosened by the difference between the actual \met\ and the minimum \met\ required in the SR,
to avoid selecting harder jets or leptons in the low-\met region.
The $\met/\meff$ requirement is loosened similarly.
For \texttt{Rpc2L0b}, the small number of events in the low-\met\ region, compared with the SR,
is due to the combined effects of the \meff\ and $\met/\meff$ requirements,
preventing high-\meff\ events from being selected.
 
\begin{figure}[!t]
\includegraphics[width=\textwidth]{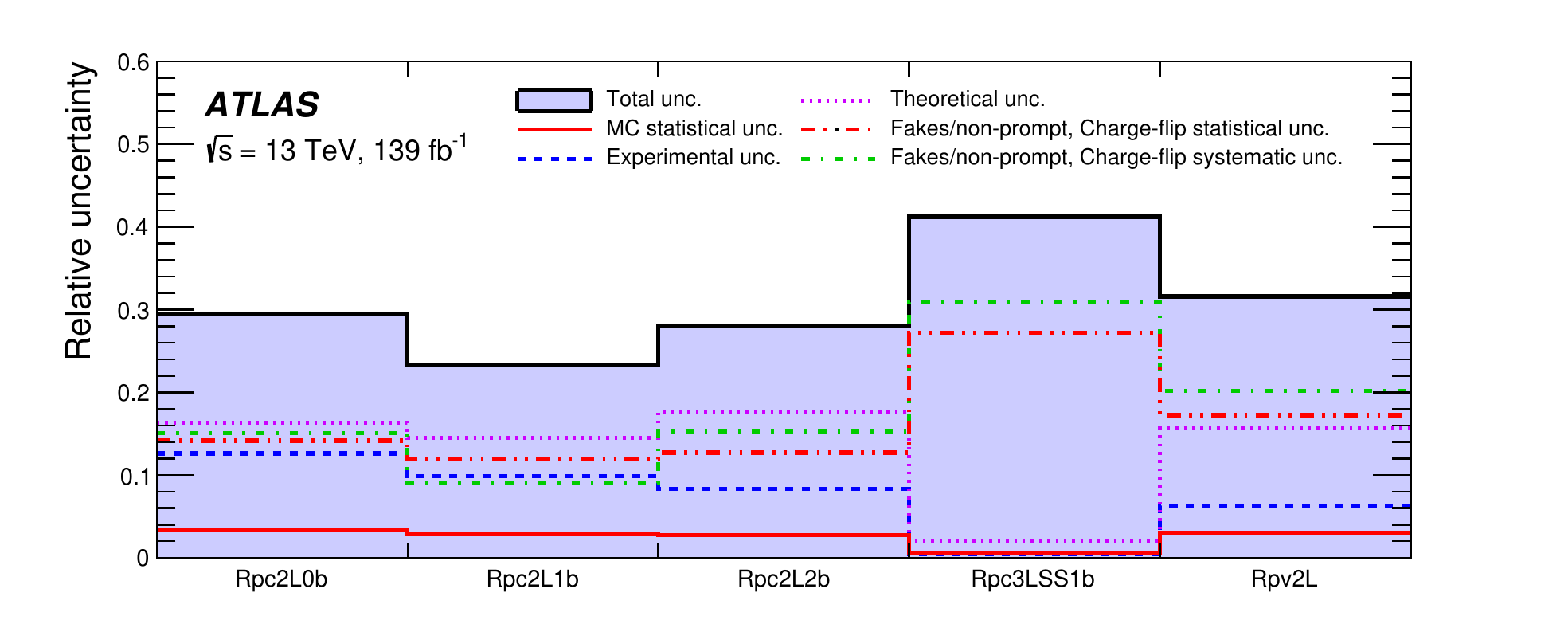}
\caption{
Contributions of different categories of uncertainties
relative to the expected background yields in the five SRs.
The statistical uncertainties
originate from the limited number of preselected or opposite-sign data events
used in the matrix method and the charge-flip electron background estimate, respectively,
as well as the effect of limited numbers of simulated events for SM processes.
}
\label{fig:systematics}
\end{figure}
 
Figure~\ref{fig:systematics} presents a summary of the contributions from different sources of systematic uncertainty
to the total uncertainties in the predicted total background yields.
These range from $23\%$ to $41\%$,
and are always smaller than the statistical uncertainties in the observed event yields.
 
\begin{table}[h]
\caption{
Computed $95\%$ CL upper limits on the numbers of BSM events $S^{95}$,
as well as the $\pm 1\sigma$ expected fluctuations around the mean expected limit.
These are also translated into upper limits on the visible cross-section $\sigma_\text{vis}$.
The $p$-values $p(s=0)$ give the probabilities to observe a deviation
from the predicted background at least as large as that in the data.
They are capped at $0.50$.
}
\label{tab:upper-limits}
 
\setlength{\tabcolsep}{0.0pc}
\begin{tabular*}{\textwidth}{@{\extracolsep{\fill}}lccccc}
\noalign{\smallskip}\hline\hline\noalign{\smallskip}
Signal region & $\sigma_\text{vis}$ [fb]  &  $S_\textrm{obs}^{95}$  & $S_\textrm{exp}^{95}$ & $p(s=0)$  \\[2pt]
\noalign{\smallskip}\hline\noalign{\smallskip}
\texttt{Rpc2L0b}    & $0.05$ &  $\phantom{1}7.6$ & $ { 6.4 }^{ +3.2 }_{ -2.0 }$  & $ 0.33$ \\[2pt]
\texttt{Rpc2L1b}    & $0.08$ &  $11.6$ & $ { 7.3 }^{ +3.6 }_{ -2.3 }$ & $ 0.09$ \\[2pt]
\texttt{Rpc2L2b}    & $0.09$ &  $12.4$ & $ { 8.7 }^{ +4.0 }_{ -2.7 }$ & $ 0.14$ \\[2pt]
\texttt{Rpc3LSS1b}  & $0.04$ &  $\phantom{1}6.2$ & $ { 5.7 }^{ +2.9 }_{ -1.8 }$ & $ 0.41$ \\[2pt]
\texttt{Rpv2L}      & $0.05$ &  $\phantom{1}6.7$ & $ { 6.9 }^{ +3.2 }_{ -2.0 }$ & $ 0.50$ \\[2pt]
\noalign{\smallskip}\hline\hline\noalign{\smallskip}
\end{tabular*}
\end{table}
 
Upper limits at $95\%$ confidence level (CL) on possible BSM contributions to the SRs
are computed with the HistFitter framework~\cite{Baak:2014wma},
relying on a profile-likelihood-ratio test~\cite{Cowan:2010js}
and following the CL$_\text{s}$ prescription~\cite{Read:2002hq}.
The hypothesis tests are performed for each of the SRs independently.
The likelihood is built as the product of a Poisson probability distribution
describing the observed number of events in the SR and the probability distributions of the nuisance parameters encoding the systematic uncertainties.
The latter are Gaussian distributions for all sources,
including statistical uncertainties
arising from the limited number of preselected or opposite-sign data events in the estimation of the reducible background,
or the limited number of simulated events.
Correlations of a given nuisance parameter between the backgrounds and the signal are taken into account when relevant.
 
Table~\ref{tab:upper-limits} presents $95\%$ CL upper limits
on the number of BSM events, $S^{95}$, that may contribute to the SRs.
Normalising these by the integrated luminosity $L$ of the data sample,
they can be interpreted as upper limits on the visible BSM cross-section ($\sigma_\text{vis}$),
defined as $\sigma_\text{vis}=\sigma_\text{prod}\times \mathcal A\times \epsilon = S^{95} / L$,
where $\sigma_\text{prod}$ is the production cross-section of an arbitrary BSM signal process,
and $\mathcal{A}$ and $\epsilon$ are the corresponding fiducial acceptance and reconstruction efficiencies for the relevant SR.
These limits are computed with asymptotic approximations of the probability distributions
of the test statistic under the different hypotheses~\cite{Cowan:2010js}.
They were confirmed to be within $10\%$ of an alternative computation based on pseudo-experiments.
The probability of the observations being compatible with the SM-only hypothesis is quantified
by the $p$-values displayed in Table~\ref{tab:upper-limits};
the smallest, for \texttt{Rpc2L1b}, corresponds to about 1.3 standard deviations.
 
\section{Exclusion limits on SUSY scenarios}
\label{sec:limits}
Exclusion limits are computed for the masses of superpartners
involved in the benchmark SUSY signal scenarios shown in Figure~\ref{fig:benchmarks},
using the same statistical tools as those described in Section~\ref{sec:results}.
The limits are obtained in the context of simplified models~\cite{Alwall:2008ve,Alwall:2008ag,Alves:2011wf}
assuming a single production process with $100\%$ branching ratio into the chosen decay mode,
and where superpartners not involved in the process are treated as decoupled.
All superpartners are assumed to decay promptly.
The expected signal contributions to the SRs
are estimated from simulated Monte Carlo samples
produced with the \MGMCatNLOV{2.2.1} generator using LO matrix elements for the signal process with up to two extra partons.
Parton shower, hadronisation and modelling of the underlying event
were performed using the \PYTHIAV{8.230} generator~\cite{Sjostrand:2014zea} with the A14 tune~\cite{ATL-PHYS-PUB-2014-021},
using the CKKW-L matching prescription~\cite{Lonnblad:2011xx}
with a matching scale set to one quarter of the mass of the gluinos or squarks produced in the interaction.
The samples were processed through a fast simulation of the ATLAS detector
using a parameterisation of the calorimeter response but \textsc{Geant}4 for the ID and MS~\cite{SOFT-2010-01, ATL-PHYS-PUB-2010-013}.
Such an approach is known to be appropriate for the standard reconstruction techniques described in Section~\ref{sec:reconstruction},
and alternative corrections and scale factors to those evoked in Sections~\ref{sec:reconstruction} and~\ref{sec:prompt_bkg} are employed.
The samples are normalised to the `NNLO$_\text{approx}$+NNLL' reference cross-sections~\cite{Beenakker:2016lwe},
which combine near-threshold approximate next-to-next-to-leading-order corrections~\cite{Beneke:2009ye} to the NLO cross-section
with a resummation of soft gluon divergences at next-to-next-to-leading-logarithm accuracy~\cite{Beenakker:2016lwe}.
Corresponding uncertainties are taken from envelopes of cross-section predictions
using different PDF sets and factorisation and renormalisation scales, as described in Ref.~\cite{Butterworth:2015oua}.
They range from $12\%$ to $20\%$ for gluino masses from 1 to \SI{2}{TeV},
and from $7\%$ to $11\%$ for top or bottom squark masses from \SI{400}{GeV} to \SI{1}{TeV}.

\begin{figure}[t!]
\begin{subfigure}{0.49\textwidth}\centering
\includegraphics[width=\textwidth]{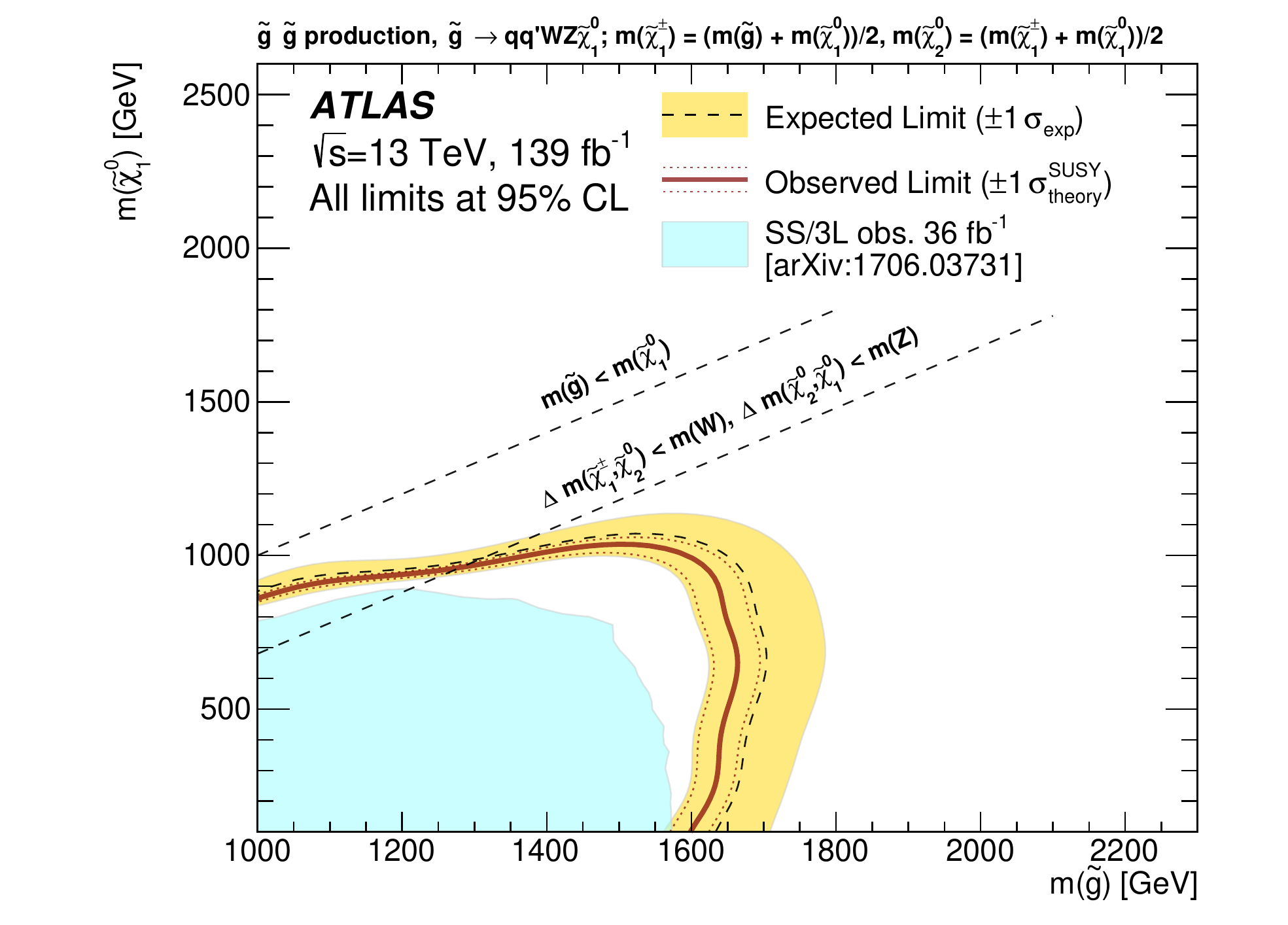}
\caption{\texttt{Rpc2L0b}: $\gluino\to qq'WZ\ninoone$}\label{fig:limits-gluino-qqWZN1}
\end{subfigure}
\begin{subfigure}{0.49\textwidth}\centering
\includegraphics[width=\textwidth]{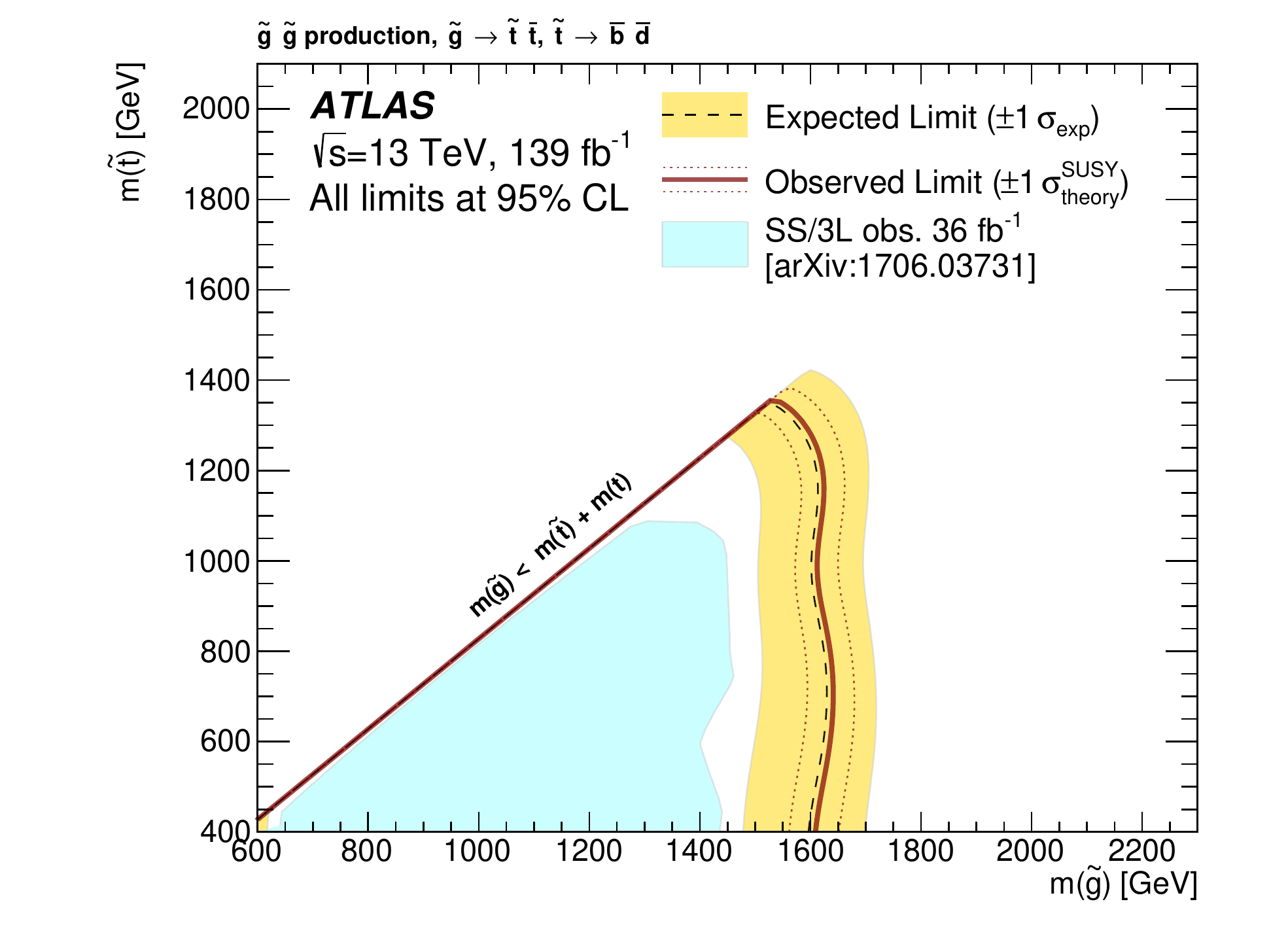}
\caption{\texttt{Rpv2L}: $\gluino\to tbd$}\label{fig:limits-gluino-tbd}
\end{subfigure}
 
\caption{
$95\%$ CL exclusion limits on the production of pairs of gluinos,
assuming production cross-sections as in Ref.~\cite{Beenakker:2016lwe}
and 100\% branching ratios into the decay modes
illustrated in Figures~\ref{fig:gluino-qqWZN1} and~\ref{fig:gluino-tbd} for the left and right plots, respectively.
The limits are determined from the expected contributions of these processes to the
\texttt{Rpc2L0b} (left) and \texttt{Rpv2L} (right) SRs.
The coloured bands display the $\pm 1\sigma$ ranges of the expected fluctuations around the mean expected limit,
in the absence of contributions from the sought-for signals.
They do not account for uncertainties in the signal process cross-sections,
the impact of which is illustrated by the dashed lines around the observed limits.
The figures show for reference the reach of the previous analysis~\cite{SUSY-2016-14}.
}
\label{fig:limits_gluinos}
\end{figure}
 
Exclusion limits on the masses of gluinos are shown in Figure~\ref{fig:limits_gluinos}.
The limits in Figure~\ref{fig:limits-gluino-qqWZN1}
are set for pair production of gluinos in an $R$-parity-conserving scenario (Figure~\ref{fig:gluino-qqWZN1}) with decoupled squarks
and gluinos decaying in two steps with intermediate $\chinoonepm$ and $\ninotwo$ into jets, weak bosons and the LSP $\ninoone$.
The $\chinoonepm$ mass is assumed to be $0.5\times\{m(\gluino)+m(\ninoone)\}$,
while the $\ninotwo$ mass is similarly $0.5\times\{m(\chinoonepm)+m(\ninoone)\}$.
The weak bosons produced in the cascade decays might be off shell,
if $\Delta m(\chinoonepm,\ninotwo)<m_W$
or $\Delta m(\ninotwo,\ninoone)<m_Z$.
The limits in Figure~\ref{fig:limits-gluino-tbd}
are set for pair production of gluinos in an $R$-parity-violating scenario (Figure~\ref{fig:gluino-tbd})
where gluinos decay via top squarks into $tbd$ or $tbs$ final states (experimentally indistinguishable)
when $\lambda''_{313}$ or $\lambda''_{323}$ couplings are non-zero.
Sensitivity to these two scenarios is provided by the SRs \texttt{Rpc2L0b} and \texttt{Rpv2L},
and allows exclusion of gluino masses below \SI{1.6}{TeV} for $\ninoone$ masses up to \SI{1}{TeV} or $\stopone$ masses up to \SI{1.2}{TeV}.
For gluino masses around the exclusion limits,
the signal $\mathcal{A}\times\epsilon$ is as large as 0.9\% for \texttt{Rpc2L0b} and 0.7\% for \texttt{Rpv2L}.
 
\begin{figure}[t!]
\begin{subfigure}{0.49\textwidth}\centering
\includegraphics[width=\textwidth]{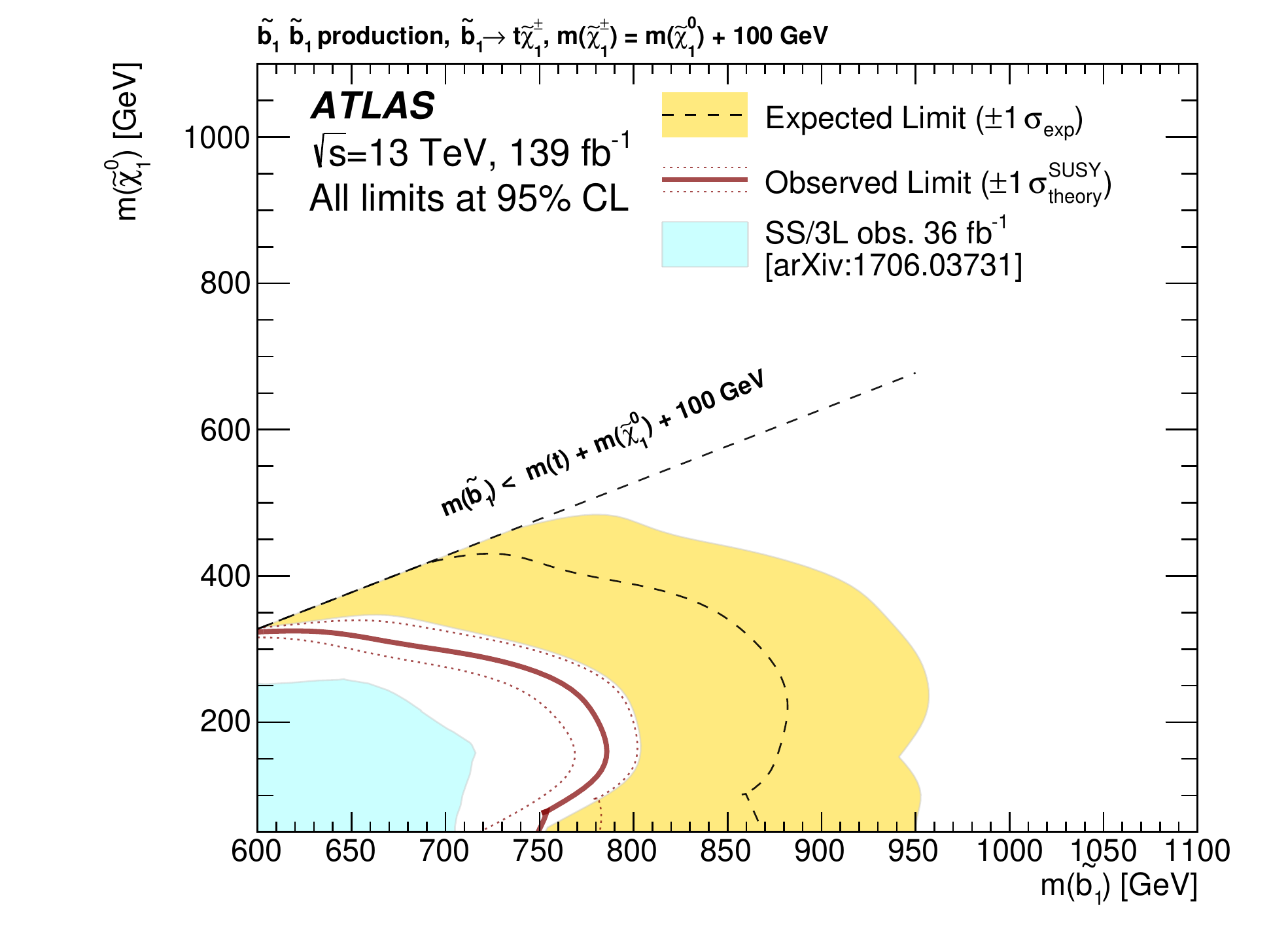}
\caption{\texttt{Rpc2L1b}+\texttt{Rpc2L2b}: $\tilde b_1\to tW^-\ninoone$}\label{fig:limits-sbottom-tWN1}
\end{subfigure}
\begin{subfigure}{0.49\textwidth}\centering
\includegraphics[width=\textwidth]{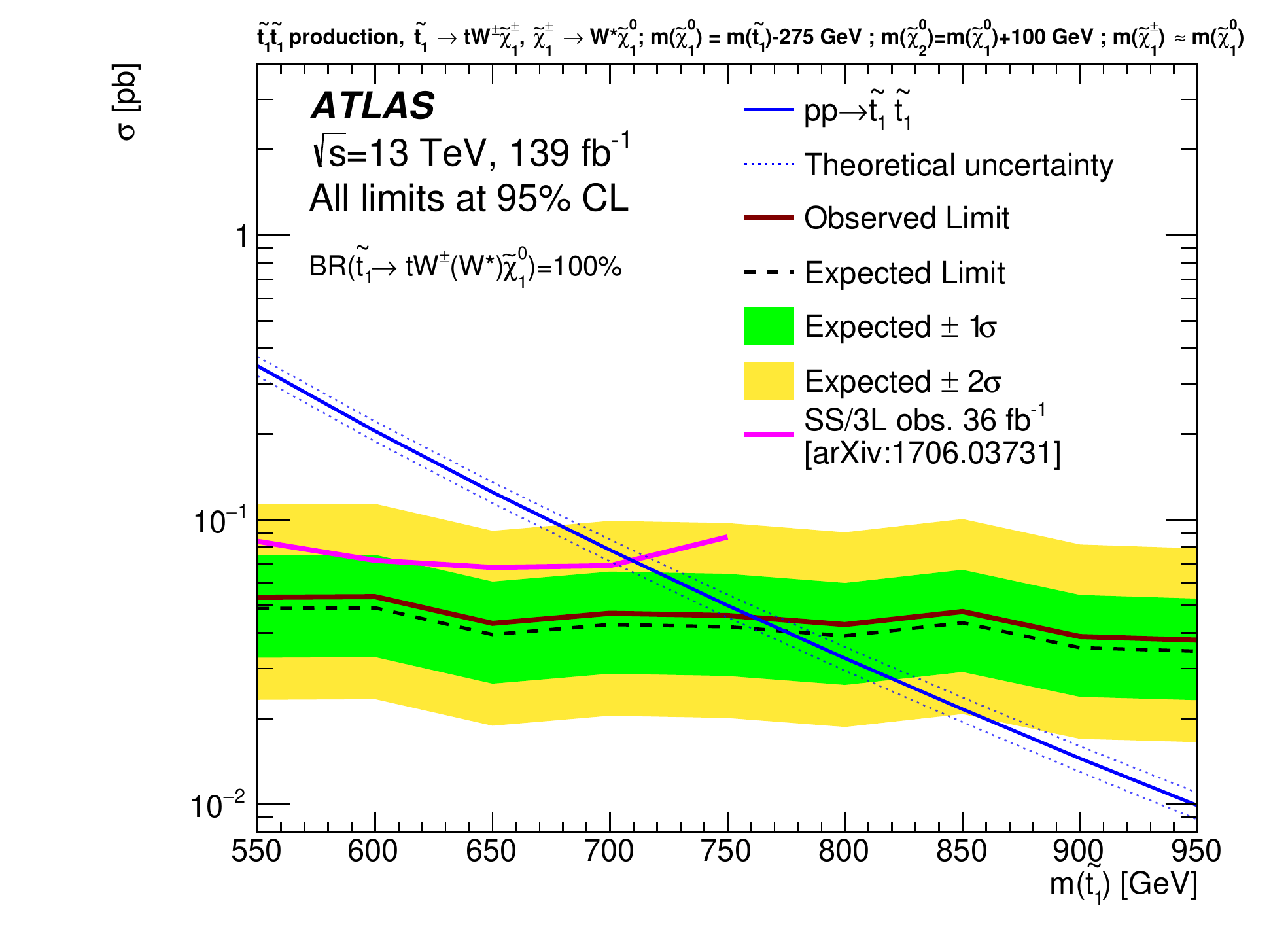}
\caption{\texttt{Rpc3LSS1b}: $\stopone\to tW^\mp\chinoonepm$}\label{fig:limits-stop-tWWN1}
\end{subfigure}
 
\caption{
$95\%$ CL exclusion limits on the production of pairs of third-generation squarks,
assuming production cross-sections as in Ref.~\cite{Beenakker:2016lwe} (for the left plot)
and 100\% branching ratios into the decay modes
illustrated in Figures~\ref{fig:sbottom-tWN1} and~\ref{fig:stop-tWWN1} for the left and right plots, respectively.
The limits are determined from the expected contributions of these processes to the
\texttt{Rpc2L1b} and \texttt{Rpc2L2b} (left) and \texttt{Rpc3LSS1b} (right) SRs.
The coloured bands display the $\pm 1\sigma$ ranges (as well as $\pm 2\sigma$ for the right figure)
of the expected fluctuations around the mean expected limit,
in the absence of contributions from the sought-for signals.
They do not account for uncertainties in the signal process cross-sections,
the impact of which is illustrated by the dashed lines around the observed limits.
The figures show for reference the reach of the previous analysis~\cite{SUSY-2016-14}.
}
\label{fig:limits_squarks}
\end{figure}

Exclusion limits on the masses of third-generation squarks are shown in Figure~\ref{fig:limits_squarks}.
The limits in Figure~\ref{fig:limits-sbottom-tWN1}
are set for pair production of bottom squarks in an $R$-parity-conserving scenario (Figure~\ref{fig:sbottom-tWN1}) with decoupled gluinos and squarks of other flavours,
with $\sbottomone$ squarks decaying via an intermediate $\chinoonepm$ into a top quark, a $W$ boson and the LSP $\ninoone$.
The mass of the charginos $\chinoonepm$ are assumed equal to $m(\ninoone)+\SI{100}{GeV}$.
For each point of the $\{m(\sbottomone),\,m(\ninoone)\}$ parameter space,
\texttt{Rpc2L1b} and \texttt{Rpc2L2b} is chosen according to which provides better expected sensitivity.
The former provides sensitivity over most of the plane, while the latter provides some complementarity in the low-$m(\ninoone)$ region.
The transition between the two regions corresponds to the breaking of the exclusion limits smoothness at $m(\ninoone)\approx\SI{100}{GeV}$.
The limits in Figure~\ref{fig:limits-stop-tWWN1}
are set for pair production of top squarks (Figure~\ref{fig:stop-tWWN1})
decaying into a top quark, a chargino and a $W$ boson via a neutralino $\ninotwo$.
The masses are chosen similarly to Ref.~\cite{Huang:2015fba}
such that $m(\stopone)=m(\ninotwo)+m_t=m(\ninoone)+\SI{275}{GeV}$ to suppress $\ninotwo\to H\ninoone$ and have $\ninotwo\to W^\mp\chinoonepm$ as the dominant decay mode,
and $m(\chinoonepm)\approx m(\ninoone)$ such that products of the former's decay into the LSP $\ninoone$ are invisible.
Upper limits on the production cross-section are provided as a function of the $\stopone$ mass;
top squarks with masses up to \SI{750}{GeV} are excluded.
For squark masses around the exclusion limit,
the signal $\mathcal{A}\times\epsilon$ is as large as 0.2\% for \texttt{Rpc2L1b} and \texttt{Rpc2L2b}, and close to 0.1\% for \texttt{Rpc3LSS1b}.
 
\FloatBarrier

\section{Conclusion}
\label{sec:conclusion}
A search for supersymmetry in events with same-sign leptons and jets is presented.
The analysis is performed with $pp$ collision data
collected at a centre-of-mass energy of $\sqrt{s}=\SI{13}{TeV}$
between 2015 and 2018 with the ATLAS detector at the LHC,
corresponding to an integrated luminosity of \SI{139}{fb^{-1}}.
Five signal regions are defined to provide sensitivity to a broad range of supersymmetric processes. 
No significant excess over the yields expected from SM processes is observed, and
model-independent limits on the cross-section of possible BSM signal contributions to the signal regions are reported.
 
The results are interpreted in the framework of four simplified models featuring pair production of gluinos or third-generation squarks.
Lower limits on particle masses are derived at $95\%$ confidence level for these models,
reaching up to \SI{1.6}{TeV} for gluinos and \SI{750}{GeV} for bottom and top squarks,
raising the exclusion limits beyond those from a previous similar search made by ATLAS on a smaller \SI{13}{TeV} dataset.

\section*{Acknowledgements}
 

We thank CERN for the very successful operation of the LHC, as well as the
support staff from our institutions without whom ATLAS could not be
operated efficiently.
 
We acknowledge the support of ANPCyT, Argentina; YerPhI, Armenia; ARC, Australia; BMWFW and FWF, Austria; ANAS, Azerbaijan; SSTC, Belarus; CNPq and FAPESP, Brazil; NSERC, NRC and CFI, Canada; CERN; CONICYT, Chile; CAS, MOST and NSFC, China; COLCIENCIAS, Colombia; MSMT CR, MPO CR and VSC CR, Czech Republic; DNRF and DNSRC, Denmark; IN2P3-CNRS and CEA-DRF/IRFU, France; SRNSFG, Georgia; BMBF, HGF and MPG, Germany; GSRT, Greece; RGC and Hong Kong SAR, China; ISF and Benoziyo Center, Israel; INFN, Italy; MEXT and JSPS, Japan; CNRST, Morocco; NWO, Netherlands; RCN, Norway; MNiSW and NCN, Poland; FCT, Portugal; MNE/IFA, Romania; MES of Russia and NRC KI, Russia Federation; JINR; MESTD, Serbia; MSSR, Slovakia; ARRS and MIZ\v{S}, Slovenia; DST/NRF, South Africa; MINECO, Spain; SRC and Wallenberg Foundation, Sweden; SERI, SNSF and Cantons of Bern and Geneva, Switzerland; MOST, Taiwan; TAEK, Turkey; STFC, United Kingdom; DOE and NSF, United States of America. In addition, individual groups and members have received support from BCKDF, CANARIE, Compute Canada and CRC, Canada; ERC, ERDF, Horizon 2020, Marie Sk{\l}odowska-Curie Actions and COST, European Union; Investissements d'Avenir Labex, Investissements d'Avenir Idex and ANR, France; DFG and AvH Foundation, Germany; Herakleitos, Thales and Aristeia programmes co-financed by EU-ESF and the Greek NSRF, Greece; BSF-NSF and GIF, Israel; CERCA Programme Generalitat de Catalunya and PROMETEO Programme Generalitat Valenciana, Spain; G\"{o}ran Gustafssons Stiftelse, Sweden; The Royal Society and Leverhulme Trust, United Kingdom.
 
The crucial computing support from all WLCG partners is acknowledged gratefully, in particular from CERN, the ATLAS Tier-1 facilities at TRIUMF (Canada), NDGF (Denmark, Norway, Sweden), CC-IN2P3 (France), KIT/GridKA (Germany), INFN-CNAF (Italy), NL-T1 (Netherlands), PIC (Spain), ASGC (Taiwan), RAL (UK) and BNL (USA), the Tier-2 facilities worldwide and large non-WLCG resource providers. Major contributors of computing resources are listed in Ref.~\cite{ATL-GEN-PUB-2016-002}.
 
 
\printbibliography
 
\clearpage
 
\begin{flushleft}
\hypersetup{urlcolor=black}
{\Large The ATLAS Collaboration}

\bigskip

\AtlasOrcid[0000-0002-6665-4934]{G.~Aad}$^\textrm{\scriptsize 102}$,    
\AtlasOrcid[0000-0002-5888-2734]{B.~Abbott}$^\textrm{\scriptsize 129}$,    
\AtlasOrcid{D.C.~Abbott}$^\textrm{\scriptsize 103}$,    
\AtlasOrcid[0000-0002-2788-3822]{A.~Abed~Abud}$^\textrm{\scriptsize 71a,71b}$,    
\AtlasOrcid[0000-0002-1002-1652]{K.~Abeling}$^\textrm{\scriptsize 53}$,    
\AtlasOrcid[0000-0002-2987-4006]{D.K.~Abhayasinghe}$^\textrm{\scriptsize 94}$,    
\AtlasOrcid[0000-0002-8496-9294]{S.H.~Abidi}$^\textrm{\scriptsize 167}$,    
\AtlasOrcid[0000-0002-8279-9324]{O.S.~AbouZeid}$^\textrm{\scriptsize 40}$,    
\AtlasOrcid{N.L.~Abraham}$^\textrm{\scriptsize 156}$,    
\AtlasOrcid[0000-0001-5329-6640]{H.~Abramowicz}$^\textrm{\scriptsize 161}$
% ...
\bigskip
\\

% ...
$^{40}$Niels Bohr Institute, University of Copenhagen, Copenhagen; Denmark.\\
% ...
$^{53}$II. Physikalisches Institut, Georg-August-Universit\"{a}t G\"ottingen, G\"ottingen; Germany.\\
% ...
$^{71}$$^{(a)}$INFN Sezione di Pavia;$^{(b)}$Dipartimento di Fisica, Universit\`a di Pavia, Pavia; Italy.\\
% ...
$^{94}$Department of Physics, Royal Holloway University of London, Egham; United Kingdom.\\
% ...
$^{102}$CPPM, Aix-Marseille Universit\'e, CNRS/IN2P3, Marseille; France.\\
$^{103}$Department of Physics, University of Massachusetts, Amherst MA; United States of America.\\
% ...
$^{129}$Homer L. Dodge Department of Physics and Astronomy, University of Oklahoma, Norman OK; United States of America.\\
% ...
$^{156}$Department of Physics and Astronomy, University of Sussex, Brighton; United Kingdom.\\
% ...
$^{161}$Raymond and Beverly Sackler School of Physics and Astronomy, Tel Aviv University, Tel Aviv; Israel.\\
% ...
$^{167}$Department of Physics, University of Toronto, Toronto ON; Canada.\\
% ...
$^{a}$ Also at Borough of Manhattan Community College, City University of New York, New York NY; United States of America.\\
$^{b}$ Also at CERN, Geneva; Switzerland.\\
% ...

\end{flushleft}


\end{document}